# Learning from Past Mistakes: AI-Assisted Analytics of Software Defects and Their Fixes for Future Defect Prevention – An Industrial Case Study

Vəhid Goruslu[1]
Queen's University Belfast, Belfast, United Kingdom
v.garousi@qub.ac.uk
Azerbaijan Technical University, Baku, Azerbaijan
vahid.garousi@aztu.edu.az
ORCID: 0000-0001-6590-7576

**Abstract:**

*Context.* Despite advances in software testing and quality assurance, defects continue to recur in industrial software systems. Defect data are routinely collected in issue-tracking systems, but their use is largely confined to detection, reporting, and short-term resolution, leaving the knowledge embedded in historical defects and their fixes under-exploited for prevention and long-term improvement.

*Objective.* This paper investigates how historical defect and defect-fix data can be analyzed, with AI assistance under human oversight, to identify the human-error origins of recurring defects and to feed that understanding back into prevention and QA improvement.

*Method.* We developed an AI-assisted, human-in-the-loop analytics approach that classifies defects by failure type and by human-error origin across the SDLC, determines how each could have been prevented, and synthesizes engineer-focused capability-building insights. It was deployed and evaluated in a large software company through an industrial case study, using practitioner surveys and interviews, classification-reliability checks, and an effort-normalized analysis of defect density.

*Results.* Practitioners and independent review judged the classifications accurate; the outputs were rated useful and actionable; and defect density decreased after the introduction of the feedback, a tentative association we interpret as consistent with, but not proof of, a positive effect, given the study's uncontrolled pre/post design.

*Conclusion.* AI-assisted, human-governed analysis of defect-fix data can help shift organizations from defect detection toward prevention, continuous learning and improvement.

[1] Alternative name spelling: Vahid Garousi

## Contents

## 1 NTRODUCTION

Despite significant advances in software testing methods and tools, defects continue to recur in industrial software systems. In many organizations, defect data are systematically collected through issue-tracking systems, yet their primary use remains limited to defect detection, reporting, and short-term resolution. As a result, the valuable information embedded in historical defects and their fixes is not often exploited for defect prevention or long-term process improvement [1]. This detection-centric use of defect data constrains an organization's ability to understand where defects originate, why similar mistakes recur, and how preventive actions can be prioritized across the software development lifecycle (SDLC).

The central principle of this work is *learning from past mistakes*: treating an organization's accumulated defects and their fixes not as isolated incidents to be closed, but as a body of evidence about the human errors that produce defects and about how those defects can be prevented. This principle is long established in software engineering—in defect causal analysis [2] and in the packaging of experience for reuse advocated by the Quality Improvement Paradigm [3] and Experience Factory [4]—yet in practice much defect data remains unanalyzed, in part because systematically mining large volumes of free-text defect and fix records is labor-intensive. Recent advances in generative AI create an opportunity to revisit this principle at scale, provided the analysis remains under appropriate human control.

This paper presents an AI-assisted, human-in-the-loop analytics approach to systematically extract insights from defect–fix data in industrial contexts. Rather than treating defects as isolated events, the approach analyzes defect origins, recurring mistake patterns, and fix recurrences to support software process improvement (SPI), software quality assurance (SQA), and developer capability building. The approach was applied in a large industrial software organization and has been integrated into its regular quality practices, illustrating how AI-assisted analytics can help shift organizational focus from defect detection toward defect prevention and continuous learning.

The contributions of this paper are fourfold:

- An AI-assisted, human-in-the-loop approach for analyzing historical defect *and* defect-fix data, which classifies defects by both their observed failure and their human-error origin across the SDLC, determines how each could have been prevented, and synthesizes the results into feedback for process improvement, quality assurance, and engineer capability building.
- A fine-grained, origin-level view of defects that localizes human errors more precisely than conventional SDLC-phase classification, and which reveals concentrations—such as back-end and data-related origins—that a coarse view obscures.
- An industrial case study evaluating the approach in a real organization, triangulating practitioner judgement, classification-reliability checks, and an effort-normalized pre/post analysis of defect density, together with an account of the approach's adoption and sustained use.
- A candid treatment of the practical and ethical considerations of engineer-level defect analysis, including the safeguards required to use it for learning rather than evaluation.

The remainder of this paper is organized as follows. Section 2 describes the research approach, the industrial context and problem, and related work. Section 3 presents the AI-assisted analytics approach and how its results were used for improvement. Section 4 reports the evaluation—its design, research questions, and results—and the approach's current use in the organization. Section 5 discusses lessons learned, implications, generalizability, and threats to validity, and Section 6 concludes.

## 2 RESEARCH APPROACH, STUDY CONTEXT, AND RELATED WORK

### 2.1 Overall research approach: Action Research within a technology transfer process

The research adopted a pragmatic action-research approach [5], involving close collaboration with practitioners to analyze existing defect data, introduce analytics-driven improvements, and reflect on outcomes in real-world settings. Action research was appropriate because the researcher (author) was embedded in the participating company as a consultant, and the approach was developed, applied, and refined through successive iterations rather than delivered as a finished artifact for external assessment.

To structure this work, we followed a published model for technology transfer in practice, specific to software engineering [6], which describes how a solution developed together with industry moves, in stages, from an initial practical problem toward released and sustained use. Its authors present the model as an instantiation offered for inspiration rather than as a rigid prescription [6], and we adapted it to the circumstances of this study, as we explain next.

Figure 1 maps our pragmatic research approach onto this transfer process and also serves as a roadmap for the paper. The work began with a concrete problem observed in practice (to be discussed in Section 2.2), positioned w.r.t. the state of the art and practice in Section 2.3, and then formulated as a research problem in Section 2.4, and. We then developed a candidate solution: the AI-assisted defect and defect-fix analytics approach described in Section 3. We then examined its usefulness through the model's static validation, in which practitioners assessed the accuracy and usefulness of the analytics outputs, and dynamic validation, in which the approach was piloted over live development activity across the company's systems. In our study, these two validation stages correspond to two iterations of applying and refining the approach in practice. Section 4 describes how they were conducted as an industrial case study, and reports their results. The final stage of the model, releasing the approach into regular organizational use, is addressed in Section 4.7.

We departed from that technology-transfer model [6] in one respect. The model includes a lab-validation stage, in which a candidate solution is first evaluated in an academic setting before industrial trials [6]. Our approach had no meaningful academic setting (pre-industrial form), since it was conceived, built, and validated inside the industrial context using that context's own defect data, so we omitted this stage. Overall, the project has run for more than 1.5 years as of this writing, and the approach has passed through several phases and iterations to ensure that it meets industry needs and be as systematic as possible. We believe it has been a successful instance of industry-academia collaboration [7-10], following the above pragmatic research approach.

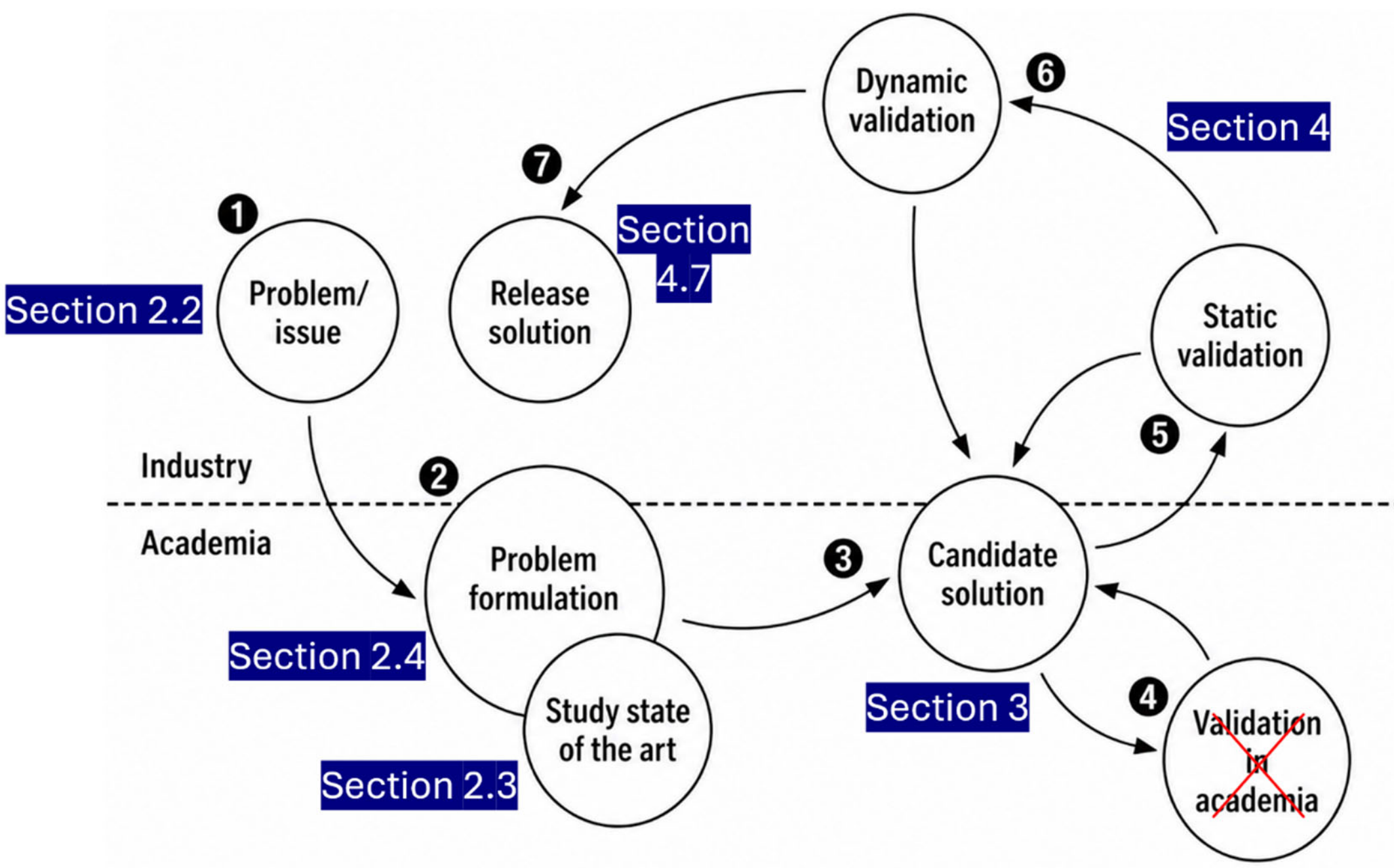

**Figure 1- Overall research approach: Action research within a technology transfer process, based on [6]**

## 2.2 Industrial context, problem and issue motivating this work

The industrial context that motivated the approach reported in this paper, and its case study (Section 4), is a large software company whose location is undisclosed to preserve anonymity. The company develops complex, web-based systems for domains including government, finance, and healthcare, where correctness, data integrity, and auditability are paramount. To preserve confidentiality, the participating organization, its systems, and the individuals involved are referred to only by anonymized identifiers, and all data used in this paper have been anonymized accordingly.

Like many organizations delivering software of this scale, the company invested considerable effort in dealing with defects. Testers spent substantial time conducting manual and automated testing to detect issues, and developers spent substantial time diagnosing and fixing the defects, across a portfolio of systems in active development and maintenance. This effort, however, was directed almost entirely at detecting and resolving individual defects as they arose. Defects were treated largely as isolated incidents to be closed, rather than as a body of evidence about where and why problems were being introduced. As a result, recurring underlying causes were rarely identified, and little of the knowledge generated by fixing one defect was carried forward to prevent similar defects in future.

This situation created both a problem and an opportunity. The problem was the absence of a systematic mechanism to identify the root causes of defects and to feed that understanding back into development practice for prevention and improvement. The opportunity lay in the data the company had already accumulated: over time, its issue-tracking systems had captured a large and growing record of resolved defects, each with a description of the observed problem and information about how it was fixed. Such historical defect and fix data are a recognized source of insight into software quality [1], yet in this setting they were seldom analyzed beyond the resolution of individual issues. The work reported here was motivated by the need to turn this under-used data into a basis for software-engineering learning and improvement: identifying the human errors and process weaknesses behind recurring defects, and using that knowledge to strengthen quality assurance (QA), improve development and testing processes, and support the continued upskilling of engineers.

## 2.3 Related work: State of the art and practice

Several strands of research and industrial practice relate to this work. These include orthogonal defect classification (ODC) [11], defect causal analysis, root cause analysis [2], defect prediction [12], recent AI-based defect classification [13], human error taxonomies in software engineering (SE) [14], usage of the Skill, Rule and Knowledge (SRK) model in SE [15-17], and Basili's Quality Improvement Paradigm [3] and Experience Factory [4].

A long-standing line of work seeks to learn from defects by classifying them and analyzing their causes. Orthogonal defect classification introduced the idea of using structured defect attributes as in-process measurements to provide feedback to developers [11], while defect causal analysis focused on identifying the underlying causes of defects so that recurring problems could be prevented rather than merely fixed [2]. Complementary work has sought to characterize the human errors behind software defects through dedicated taxonomies [14], and cognitive models such as the Skill–Rule–Knowledge framework have been applied to reason about the human modes and error mechanisms involved in SE work [15-17]. Our approach shares the prevention-oriented and human-error–focused spirit of these lines of work, but differs in exploiting both defect and defect-fix data together and in doing so with AI assistance under human oversight.

A second, largely separate strand focuses on defect detection, classification, and reporting, often to improve triage or to prioritize testing effort. Predicting bugs from history demonstrated that historical defect and change data can be used to identify fault-prone components, primarily to guide testing and inspection rather than to support learning from how defects are fixed [1]. More recent AI-based approaches, such as large language model (LLM)–based bug report classification frameworks [18, 19], have improved the automation and accuracy of defect categorization, again emphasizing efficient detection and reporting over prevention. Recent studies have also applied pre-trained language models to automatic issue classification, examining factors such as the effect of data quality on classification performance [20]. Beyond classification, LLMs have been applied to root-cause analysis of operational failures; for example, Chen et al. use an LLM-based system to analyze cloud-incident data and predict root-cause categories in an industrial deployment [21]. Other studies have used LLMs to generate root-cause explanations from bug-report text to support defect understanding [22]. Our work differs in analyzing both defects and their fixes to enable learning-oriented feedback and defect prevention, rather than explanation alone. Software defect prediction more broadly has been the subject of extensive study [12], but again with the goal of anticipating where defects will occur rather than learning from how they were resolved.

Industrial experience with AI-based defect analytics illustrates the same trend. A reported industrial effort at Siemens, for example, highlights the potential of analytics and AI to monitor defect data at scale, but concentrates on defect tracking and visualization, with limited emphasis on systematically exploiting defect-fix data for organizational learning or engineer capability building [23]. The same emphasis is visible in commercially available tooling. In our systematic review of AI-powered software testing tools available on the market up to 2024 [24, 25], of all the 56 tools surveyed, a subset performed some form of analysis on test-result and failure data such as bug reports; these typically cluster or summarize failures to reduce the effort of triage. For instance, a number of those AI-powered software testing tool cluster or summarize failures to reduce triage effort—for instance, a tool such as *CloudBees Smart Tests*[1] groups failed test cases into underlying issues and summarizes logs—rather than mining fix data to support prevention and learning.

The work most closely aligned with ours in intent is Basili's Quality Improvement Paradigm and its associated Experience Factory concept, which advocate systematically packaging experience and knowledge gained from past projects so that it can be reused to improve future work [3]. Our approach can be understood as an instance of this philosophy realized with contemporary generative AI: it treats accumulated defect and defect-fix data as an experience base, uses AI-assisted analytics under human oversight to distill recurring error patterns from it, and feeds the resulting knowledge back into process improvement, quality assurance, and engineer upskilling. Framed this way, the contribution of this paper lies less in any single analytic technique than in how such analytics are integrated, in a human-governed way, into the routine quality practices of an industrial organization.

[1] cloudbees.com/capabilities/cloudbees-smart-tests

It is worth being explicit about what is and is not novel in this work. The individual building blocks are largely established: defect and human-error classification draw on the orthogonal-defect-classification and human-error-taxonomy traditions [14], LLM-based defect classification has been demonstrated elsewhere [18-21], and the idea of packaging past experience for reuse is the core of the Quality Improvement Paradigm and Experience Factory [3, 4]. The contribution of this paper is therefore not a new algorithm or classification technique. Rather, it is the integration of these elements into a single, human-governed AI-assisted pipeline that analyzes defect *fixes* as well as defects, attributes them to fine-grained human-error origins across the SDLC, and feeds the resulting insights into three organizational streams—process improvement, quality assurance, and engineer capability building—together with an account of its adoption and a longitudinal evaluation in a real industrial setting. To our knowledge, this particular integration, and its industrial evaluation, has not previously been reported.

### 2.4 Problem formulation and objectives

The industrial need described in Section 2.2 and the state of the art reviewed in Section 2.3 together define the problem addressed in this paper. On the one hand, the participating organization accumulated a large and growing record of resolved defects and their fixes, but used it almost entirely to close individual issues rather than to learn from them. On the other hand, existing research and tooling that could help are largely oriented toward detecting, predicting, or triaging defects, and comparatively little attention has been paid to systematically exploiting defect-fix data, under human oversight, to drive prevention, process improvement, and engineer learning within a single organization. Although this work is motivated by the needs of one large software company, the underlying challenge is not specific to that organization: many companies accumulate substantial defect and fix records while making limited systematic use of them. The challenge this work addresses, and the potential benefits of addressing it, are therefore likely to be relevant to a broad range of software organizations, an issue we return to when discussing generalizability in Section 5.4.

Accordingly, this work addresses the following problem: how historical defect and defect-fix data, analyzed with AI assistance under human oversight, can be used to identify the human-error sources behind recurring defects and to feed that understanding back into prevention and improvement in an industrial setting. This problem is pursued through three objectives:

1. to design an AI-assisted analytics approach that operates on both defect and defect-fix data while keeping analytical judgment with human experts;
2. to apply this approach in a real industrial context in order to characterize where defects originate, how they recur, and how they might have been prevented;
3. and to use the resulting insights to inform software process improvement, quality assurance, and the upskilling of engineers.

## 3 The Approach: AI-assisted analytics of defect and defect-fix data

This section describes the approach developed in response to the problem formulated in Section 2.4. It begins with an overview of the approach and the intuition behind it (Section 3.1), followed by a description of its inputs—the historical defect and defect-fix data it operates on (Section 3.2). Section 3.3 explains how we determined which types of analysis would yield beneficial insights, and Section 3.4 details the AI-assisted analyses themselves, including the way generative AI was used under human oversight. Section 3.5 then describes how the results of these analyses were used to drive improvements in software process, quality assurance, and engineer capability, and Section 3.6 discusses the ethical considerations and safeguards that governed the use of developer-level data.

### 3.1 Overview of the approach

Before describing the approach, we clarify the terminology used throughout this paper. We follow the established software engineering definitions [26], in which a defect arises through a chain that begins with a human *error*—a mistake made by a person during an activity such as requirements analysis, design, or coding. A human error introduces a *fault* (also termed a defect or bug): a flaw in a software artifact such as source code, a database object, or a configuration. When a fault is activated during execution, it produces a *failure*: an observable deviation of the system's behavior from what is expected. The two analyses at the core of our approach correspond to the two ends of this chain. Analysis 1 (Section 3.4.2) classifies each defect by its *failure*—the wrong behavior as observed by a user or tester—whereas Analysis 2 (Section 3.4.3) classifies each defect by its *origin*: the human error that caused it and the SDLC phase in which that error was made. In other words, Analysis 1 asks *how the system failed*, and Analysis 2 asks *what human error led to that failure and where it occurred*.

We began by conducting an in-depth analysis of the organization's historical defect and defect-fix data. While much of the existing literature analyzes defects primarily at the level of major SDLC phases, such as requirements, design, implementation, and testing [11], our initial examination revealed that this level of abstraction was insufficient for understanding the true distribution of defects in practice. We observed that defects originated not only from production code but also from test artifacts, database design and data-related decisions, configurations, deployment, and maintenance activities. This motivated a finer-grained view of defect origins, in which each defect is traced to the specific practice, process, or artifact in which it arose, rather than to a coarse development phase alone.

Figure 2 presents this view together with an overview of the overall approach. The upper part of the figure makes a point that is central to this work: defects do not originate only in source code, but can arise across the full range of software engineering practices, processes, and artifacts spanning the software development lifecycle. While defects in code are the most familiar and widely studied, we observed in the company under study that defects, traceable to human errors, occurred in every one of these artifacts and activities. Because several of these origins are less commonly discussed in the literature, we briefly illustrate each with the kind of example encountered in this study.

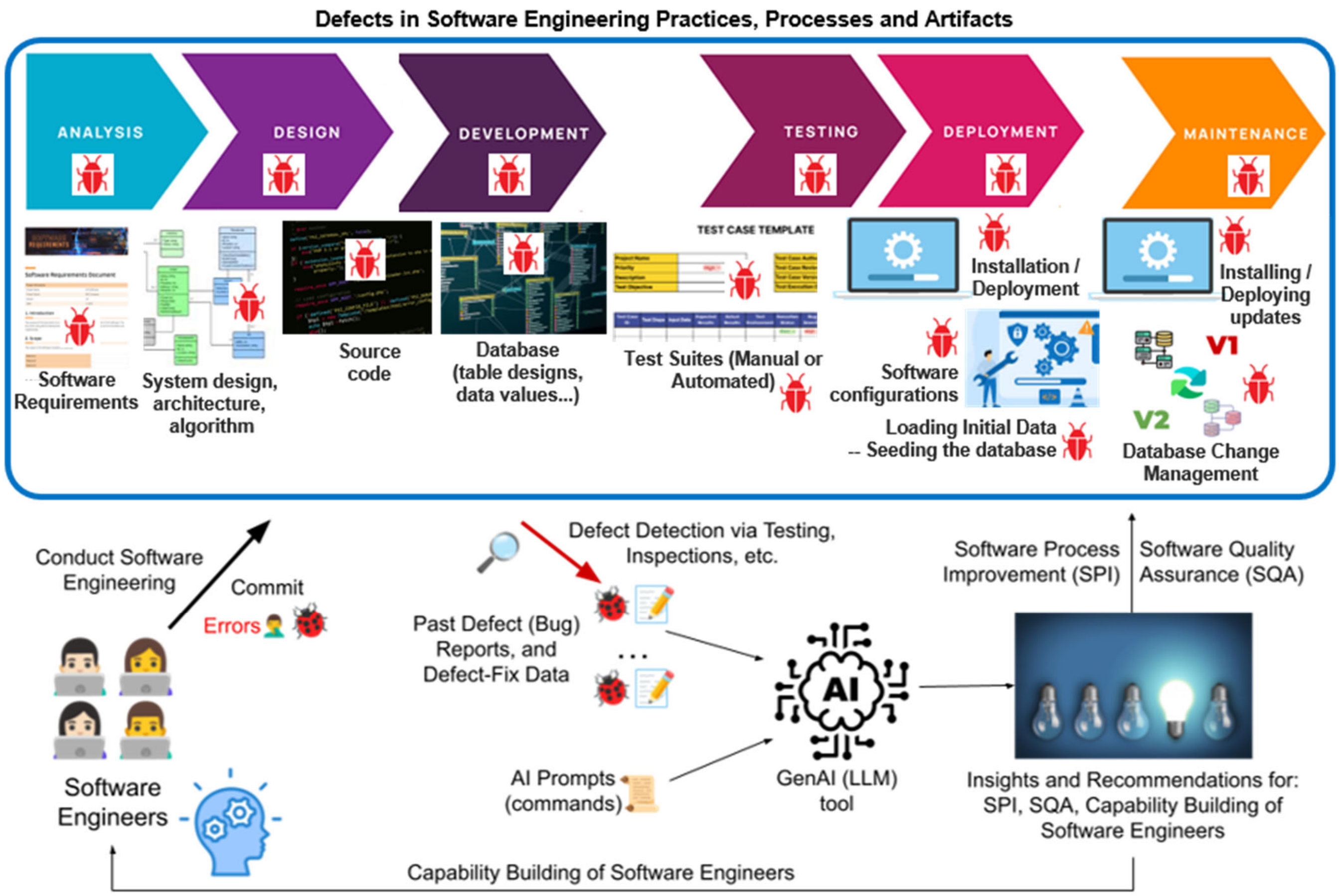

**Figure 2-A context / activity diagram showing an overview of the approach**

We observed in past defect data in the industry context that, similar to discussions in the literature (both academic and grey literature), defects can originate in *requirements*, where a misunderstood, incomplete, or incorrectly specified requirement leads to software that behaves as written but not as needed—for example, an eligibility rule for a financial calculation that was specified incorrectly and therefore implemented incorrectly. They can originate in *design*, including architectural and, notably, *database design*: a poorly designed table structure or an incorrect relationship in the data model can produce defects that surface far downstream, such as wrong results from a query that joins the affected tables. Defects in *source code* are the obvious and most frequent case, spanning front-end, back-end control logic, and database (e.g., T-SQL) code. Defects can also arise in *test artifacts* themselves: a faulty test case, an incorrect expected result, or a flawed test script can allow real defects to pass undetected or raise false alarms, so that a "defect" is in fact an fault in the test case rather than in the system under test itself.

Beyond development, several operational activities are their own sources of defects, and these are the origins most often overlooked. In *deployment*, a correct build can fail in production because a deployment step was performed incorrectly or in the wrong order. In *installation*, a required server component or tool may be missing or misinstalled—for instance, a database engine installed without a full-text search component that the application depends on, causing a feature to fail even though the application code is correct. In *configuration*, a single setting given the wrong value can disable or distort behavior while the code itself is faultless, such as a display flag initialized incorrectly so that a screen element never appears. In *loading initial (seed) data*, a system deployed with missing or incorrect reference data will raise errors when it looks up records that were never populated, even though nothing is wrong with the logic that performs the lookup. In *software maintenance*, changes made to fix one defect can introduce another—a regression—so that the act of maintenance is itself a source of defects. And in *database change management*, an incorrectly applied schema change, migration, or data update can corrupt data or duplicate records, for example when a change causes the same records to be counted or joined twice. In each of these cases, the observable failure may look similar to a coding failure, yet its true origin—and therefore its prevention—lies elsewhere.

This breadth of origins is precisely why the finer-grained, origin-level view described later (Analysis 2, Section 3.4.3) is necessary: a defect's symptom rarely reveals which of these artifacts or activities it came from, and only by tracing each defect back to its originating human error can prevention be directed to the right place.

The lower part of Figure 2 depicts the approach as a process. Software engineers, in the course of their work, occasionally introduce errors that later surface as defects; these defects and their fixes are recorded and detected through testing, inspections, and related activities; and the resulting historical defect and defect-fix data are then analyzed with the assistance of a generative AI (large language model) tool, guided by human-authored prompts and human oversight. The output of this analysis is a set of insights and recommendations that feed three streams of improvement—software process improvement, software quality assurance, and the capability building of software engineers—with the last forming a feedback loop back to the engineers whose work generated the data.

Once this comprehensive understanding of defect distribution was established, we designed the AI-assisted analytics approach around the principle of learning from past defects and their fixes. The approach systematically analyzes historical defect–fix data to extract patterns, insights, and feedback that inform software process improvement, software quality assurance, and the continued upskilling of software engineers. A distinguishing characteristic of the approach, elaborated in the subsections that follow, is that it analyzes not only defects but also the fixes applied to them, and that all analytical judgments produced with AI assistance are reviewed and validated by human experts before being acted upon.

### 3.2 Inputs: Historical defect and defect-fix data

The approach relies on historical defect reports and their associated defect-fix information, collected as part of routine development and maintenance work. The organization manages issues and tasks using the Microsoft Team Foundation Server (TFS) platform, customized to its own needs with additional fields. Each bug report contains fields conventional in the field [27], such as severity, priority, the affected system, the environment and operating system in which the issue was observed, and steps to reproduce.

Beyond this basic metadata, three textual fields are central to our analysis. The first is a bug (failure) description: a short summary of the observed problem, together with steps to reproduce it and any other useful information. This field is filled by whoever reports the defect, which may be a tester performing manual or automated testing, a business analyst exploring the system, or a staff member recording an issue reported by a client through a phone call or email. The remaining two fields are completed by the engineer who fixes the defect, after the fix: the root cause of the defect, and how the defect was resolved. Table 1 shows four example records, each pairing the reported failure with the root cause and the corrective action recorded after diagnosis and fixing.

Another field in the issue management system of the company records the SDLC phase in which the defect originated, i.e., human error type—for example, requirements, design, frontend code, backend code, database logic, configuration, or deployment. The handling of this field changed over the course of the project, and the change is itself part of the approach. Before this work, the bug-reporting form did not capture defect origin at all; in our initial analysis we therefore used AI assistance to infer the originating phase from the two existing textual fields, the root cause and the resolution description. The values of that field for the four example defects are shown in Table 1.

During the project, we added an explicit drop-down field for the originating phase, to be selected by the engineer who resolves the defect. We could have continued to rely on AI-based inference alone; we chose to add the field deliberately, to raise awareness within the organization, so that engineers would explicitly recognize the SDLC phases in which their errors were made and defects originated. Manual selection is also more reliable, because defect origin often depends on contextual knowledge and design intent that cannot be inferred automatically with confidence.

The practice of recording these two textual fields, the root cause and the resolution, had in fact been in place for several years, and the organization was already aware of their value in principle. In practice, however, a large volume of such data—several thousand records—had accumulated without being analyzed systematically. What analysis did occur was anecdotal and ad hoc, and the entries themselves were often written very briefly and were sometimes uninformative. During this project, we established a working culture of completing this information more precisely by all staff members and in more technical detail. Recording the originating phase, the root cause, and the fix description in this way was not effort-intensive: engineers reported that it typically took only a few minutes per defect, since their memory of the defect and its fix was fresh, just after completing the defect-fix task. The activity was also perceived as a useful analytical and learning exercise, prompting engineers to reflect explicitly on why each defect had occurred.

All textual data in these fields were in Azerbaijani, the working language of the organization and its engineers. For analysis and reporting in this paper, they were translated into English using Google Translate, and all translations were checked by the author, who is fluent in both languages.

**Table 1-Several example records from the dataset: defect (bug reports) and defect-fix data**

| # | Bug (failure) description (from bug report) | Written after the defect was diagnosed and fixed | | |
|---|---|---|---|---|
| | | **Root cause of the defect** | **How the defect was fixed** | **Human error type (SDLC phase)** |
| 1 | When duplicating applications in the X page, the error message "Duplicate application status not found" is encountered | We found that the “Duplicate Request” row / record does not exist in the database table related to the status, i.e. the database data was not loaded correctly (or was not set up correctly). | The "DUPLICATE" row was added to the DB table related to STATUS. | Installation error-initial data |
| 2 | Notification (workboard) in the top-right side of the screen does not work | In the configuration file, variable *notification_shown* was initialized to 1, but it should have been set to 0 (meaning that it is not shown yet, and shall be shown when needed). Note: there was no problem in the code per se; the system was configured incorrectly | In the installed configuration file, the statement *notification_shown*=0 was corrected | Code deployment error |
| 3 | After adding new information to the system, the user wants to find that information in the search window.<br><br>Technical error message: Full-Text Search is not installed, or a full-text component cannot be loaded | Incorrect installation of MS-SQL. MS-SQL 2019 Developer Edition is installed.<br>The "Full-Text and Semantic Extractions for Search"" component was not installed." | 1. Open the SQL Server installation disk or .iso file<br>2. Run the "Setup.exe" file<br>3. Select "New SQL Server stand-alone installation or add features to an existing installation"<br>4. Install the Full-Text and Semantic Extractions for Search - component | Installation error-Server/tools |
| 4 | Multiple document and numbering error when creating and copying a document in the Funds - Advance report - Subordinate person section. --- When a new document is created or copied, the system creates two copies of one document. | Code error - After creating a document, the user screen is still in the "New Document" form. The form or screen is not automatically reset or closed. As a result, when the user clicks the "Save" button again with the same data, the backend re-accepts the same data and creates a new document. That is, as a result of repeating the same operation twice, more than one document is created with the same data. | The code has been updated. After clicking the "Save" button, the user is automatically redirected to the editing screen of the newly created document. In this way, instead of staying on the same form, the user can continue on the created document and the problem of creating a new document again is eliminated. | Coding error-Frontend |

### 3.3 Determining the types of analyses that can provide beneficial insights

Given the input data described in Section 3.2, we explored which types of analysis could realistically be performed with AI assistance and would yield insights aligned with our study objectives (Section 2.4). The choice of analyses was therefore partly opportunistic: rather than beginning from a fixed analytical design, we examined what the available defect and defect-fix data could support, and selected those analyses that combined feasibility with practical value for prevention, quality assurance, and engineer learning. This is consistent with the action-research and technology-transfer setting of the work (Section 2.1), in which the approach was shaped iteratively in response to the organization's data and needs rather than specified in advance.

This exploration led us to four complementary analyses, which together move from the observable symptoms of defects toward their underlying causes and, ultimately, toward prevention and learning. Briefly, these are: classifying the type of each bug or failure as observed (Analysis 1); classifying the source of each defect in terms of the human error and SDLC phase from which it originated (Analysis 2); determining how each defect could have been prevented (Analysis 3); and extracting and synthesizing engineer-focused insights from recurring patterns across the data (Analysis 4). The first two characterize defects from two distinct angles—what was observed versus why it occurred—while the latter two turn that characterization into actionable guidance.

Before presenting these analyses, a terminological clarification is warranted. Throughout this paper, we use "analysis" (plural "analyses") to refer to a single, self-contained examination of a specific kind applied to the data—for example, Analysis 1, the classification of bug types. We use "analytics" to refer to the overall, systematic activity of deriving insights from the defect and defect-fix data, of which the individual analyses are the constituent parts. In other words, the AI-assisted analytics reported in this paper are realized through the set of four analyses described below. Section 3.4 presents each of the four analyses in turn, together with the common way in which generative AI was used, under human oversight, to carry them out. Section 3.5 then describes how the results of these analyses were used to inform the three streams of improvement—software process improvement, software quality assurance, and the capability building of engineers.

## 3.4 AI-assisted analyses

### 3.4.1 Common AI-usage approach across the analyses

The four analyses described in the following subsections differ in their inputs and objectives, but they share a common way of using generative AI and, importantly, a common discipline of human oversight. Because this shared approach applies to all four analyses, we describe it once here rather than repeating it for each. Two aspects are relevant: the protocol governing how AI was used and how its outputs were validated (Section 3.4.1.1), and the design of the prompts through which the analyses were carried out (Section 3.4.1.2).

#### 3.4.1.1 AI Usage Protocol and Human Oversight

All AI-assisted analyses were performed with the assistance of large language models, used as an analytical aid under human oversight as described below. The investigation reported in this paper began in January 2025 and has, as of this writing (July 2026), continued for more than one and a half years; the AI-assisted analyses span this period.

The specific models used evolved over the course of the work. The initial analyses, in early 2025, used OpenAI's GPT-4o, the latest GPT model available at that time. In April 2025, following internal comparisons within the company, we switched to Anthropic's Claude Sonnet for this project as well as for other coding and analytics tasks in the organization. From April 2025 onward, all analyses reported here were performed using Claude Sonnet, accessed through its desktop application. Over this period, the Claude Sonnet model itself was updated through several successive versions, and we always selected to choose its latest version available at any given time. For the purposes of our analysis, we did not observe major noteworthy differences in the quality of the model's output across these versions for our specific tasks, and the human-oversight protocol described below was applied uniformly regardless of the model version in use.

The analyses followed a consistent processing sequence: (1) the relevant defect fields were exported from the issue-tracking system; (2) a batch of records was submitted to the model with a task-specific prompt (Appendix D); (3) the model's structured, row-level output was returned with each record paired to its original text; (4) the author, with the senior technical managers where relevant, reviewed and corrected the output; and (5) validated results were recorded. Models were accessed through their standard chat interfaces (the ChatGPT web interface and, subsequently, the Claude desktop application) using default settings; we did not tune decoding parameters (e.g., temperature) or employ any custom context-management strategy beyond submitting records in batches sized to fit the model's context window.

The role of the AI in this work was that of an analytical aid, not an autonomous decision-maker. In every analysis, the AI was used to process, classify, aggregate, or summarize defect and defect-fix data according to explicit human-authored instructions, and all of its outputs were subject to human inspection and validation before being used or acted upon. The author, who is fluent in both the working language of the data (Azerbaijani) and English and is familiar with the systems concerned, reviewed the AI's outputs for correctness, corrected or discarded erroneous results, and retained final responsibility for all analytical judgments. Where an analysis involved assigning categories to defects (Analyses 1 and 2), the reliability of the resulting classification was additionally assessed, as reported in Section 4.5. This human-in-the-loop arrangement was adopted deliberately: defect analysis depends on contextual and design knowledge that current models cannot reliably supply on their own, and unvalidated AI output was not considered a trustworthy basis for organizational decisions.

To preserve confidentiality, data submitted to the AI were limited to the textual defect fields required for each analysis (for example, defect descriptions, root-cause explanations, and fix descriptions), with direct identifiers of individuals and clients removed or anonymized beforehand, consistent with the safeguards described in Section 3.6.

The human-in-the-loop component—prompt design and refinement, review and correction of AI outputs, and validation of classifications—required approximately 100-150 person-hours over the study period, performed primarily by the author, with the senior technical managers contributing to the validation. The largest share of this effort was the review and correction of AI-generated classifications across the 374 records and the analyses applied to them; prompt design and refinement, being pragmatic and iterative, accounted for a smaller portion. No costs beyond standard commercial AI subscriptions were incurred.

#### 3.4.1.2 AI Prompts and Their Engineering

We did not find it necessary to employ elaborate prompt-engineering templates or formal prompting patterns for this work. The prompts we used were deliberately pragmatic: they described the task in plain language, supplied the relevant defect fields, and specified the categories or output format required. Prompts were then refined iteratively in response to what the analysis revealed and to the organization's evolving needs, rather than being fixed in advance. A recurring and important element of this iterative refinement was the sharpening of category definitions to remove overlap: before applying a classification to the full dataset, we typically asked the model to confirm the agreed category set and to classify a small sample first, so that the logic could be checked and the wording adjusted. A representative, anonymized subset of the prompts used across all the analyses is provided in Appendix D.

Two aspects of this prompting practice are worth noting, as they supported the human-oversight protocol described in Section 3.4.1.1. First, prompts consistently requested structured, row-level output in which each defect remained paired with its original text, so that every AI-produced assignment could be checked against the underlying record. Second, the model was frequently asked to classify a small sample and expose its reasoning before full application, which allowed classification logic and category boundaries to be validated by the author before being applied to the whole dataset.

### 3.4.2 Analysis #1 - Classification of bugs (failures) types

The first analysis classifies each defect according to the type of failure observed—that is, the way in which the system behaved incorrectly from the user's or tester's point of view. Classifying failures in this way is useful for two reasons. First, it provides a concise, organization-wide picture of how systems most commonly fail, which is difficult to obtain when defects are handled only as individual issues. Second, the distribution of failure types offers an initial, symptom-level signal about where quality-assurance attention may be most needed, complementing the origin-level analysis that follows analysis #2. It is important to emphasize that this analysis classifies the symptom—the observable wrong behavior—rather than the underlying cause of the defect. The causes, expressed as the human errors and SDLC phases from which defects originated, are the subject of Analysis 2 (Section 3.4.3); the two analyses deliberately represent two distinct views of the same defects.

The classification scheme was developed jointly by the author and two senior technical managers from the company, together with the AI. Drawing on failure and defect-type distinctions established in the literature, most notably the orthogonal defect classification tradition [11], the author and the senior technical managers outlined an initial set of failure types; the AI then proposed a candidate scheme and, on request, classified small samples so that the category boundaries could be tested; and the resulting categories were refined iteratively, in consultation with the senior technical managers, to remove overlap and ambiguity. A guiding requirement throughout was orthogonality: the categories were designed to be, as far as possible, mutually exclusive. For defects that could superficially fit more than one category, a simple precedence rule was applied—if the system terminated or raised a runtime error, the defect was classified as an unhandled exception or crash; otherwise, if the intended action produced no effect at all, it was classified as a function-not-executing failure; otherwise it was assigned to the specific wrong-behavior category that best described the observed symptom. Applying this rule made the classification reproducible rather than dependent on case-by-case judgement, and it converged on the eleven failure categories shown in Figure 3.

Figure 3 presents the distribution of the 374 defects across these eleven failure types. Several observations stand out. The most frequent single failure type is unhandled exceptions or crashes (56 defects, about 15%), meaning that the most common way these systems failed, as observed, was by raising a runtime error or terminating unexpectedly rather than by failing silently. Beyond this, failures are spread broadly rather than concentrated: the next six categories each account for a similar share—Data Retrieval & Search (48), Input Validation & Handling (46), Incorrect Computation & Business Logic (43), Data Persistence & Loss (41), Workflow & State Transition (41), and Incorrect Data Display & Formatting (39)—so that no single functional area dominates. A notable pattern within this spread is the weight of data-centric failures: the four categories concerned with retrieving, persisting, displaying, and reconciling data together account for roughly 39% of all defects, indicating that a large share of failures manifest as data being wrong, missing, inconsistent, or incorrectly presented. Input-validation and business-logic failures are also prominent. At the other end, authorization failures (13) and performance problems (7) are comparatively rare in this dataset. Because this is a symptom-level view, these observations do not by themselves indicate where the defects originated—an unhandled exception, for instance, may arise from a coding error, a database-design decision, or a configuration mistake—which is precisely why the origin-level analysis in Section 3.4.3 is needed to complement it.

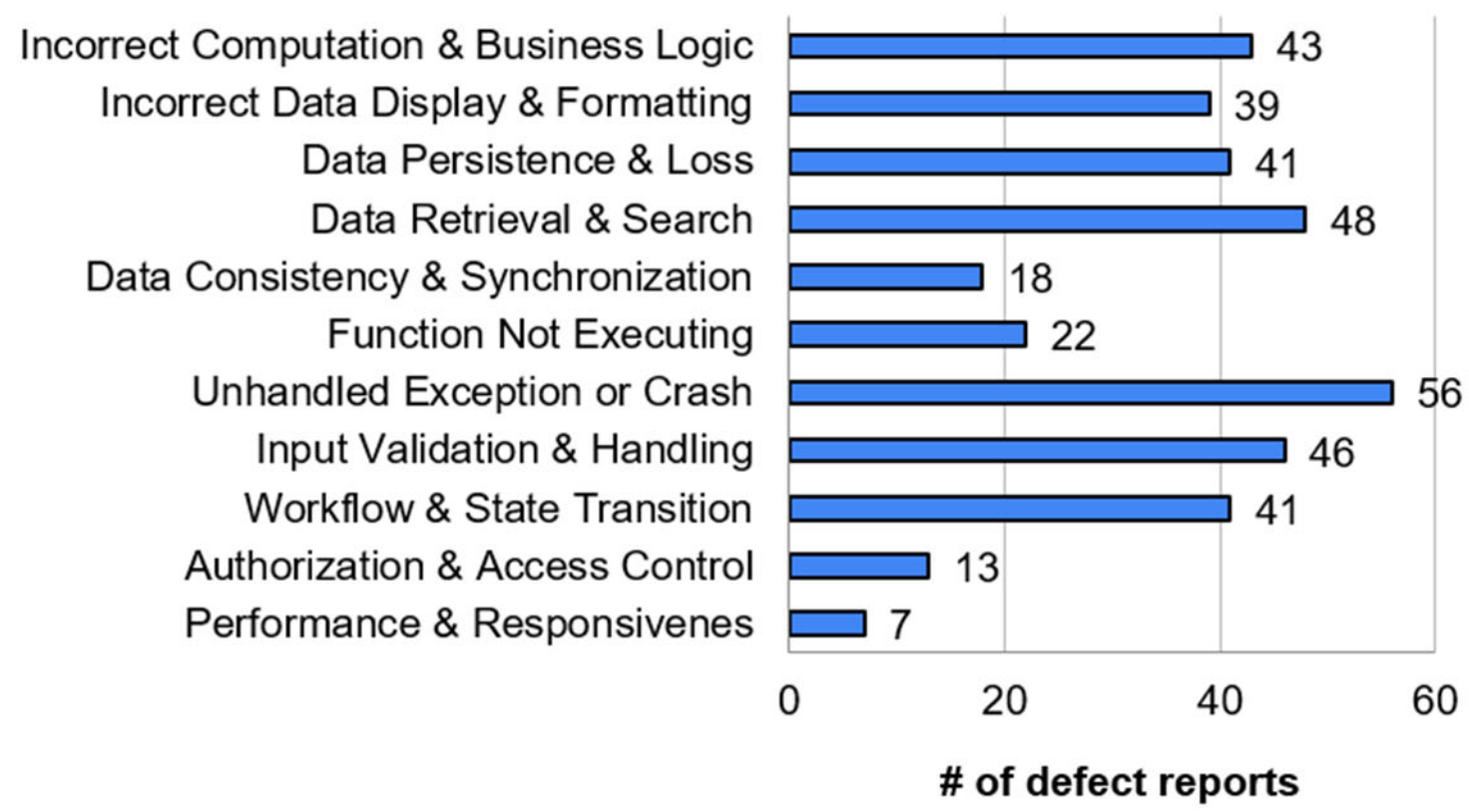

**Figure 3-Classification of bug (failure) types in the dataset**

### 3.4.3 Analysis #2 - Classification sources of defects across the SDLC: Types of Human errors leading to defects

#### 3.4.3.1 Rationale and expected benefits of the analysis

Whereas Analysis 1 characterizes defects by their observed failure, Analysis 2 characterizes them by their origin: the human error that led to each defect and the SDLC phase in which that error was made (the terminology follows the error–fault–failure chain defined in Section 3.1). This origin-level view is the more consequential of the two for prevention. Knowing how systems fail indicates where to strengthen detection, but knowing where and how the human errors behind those failures are introduced indicates where to intervene earlier, so that defects are avoided rather than merely caught. It is this shift—from the symptom of a defect to the point at which it entered the software—that connects the analytics to process improvement, quality assurance, and engineer learning (Section 3.5).

We also expected this analysis to test a premise motivating the whole approach: that classifying defect origins only at the level of coarse SDLC phases is too blunt to be actionable (Section 3.1). A finer-grained scheme, distinguishing for example between front-end coding, back-end controller logic, and back-end database (T-SQL) code within what would conventionally be a single "implementation" phase, was expected to reveal actionable concentrations that a phase-level view would obscure.

The classification scheme used for this analysis is shown in Figure 4. It comprises eighteen origin categories, which we grouped into higher-level bands reflecting the SDLC stage in which the underlying human error occurs: design (requirements analysis, database design, and UI design errors); coding (front-end, back-end controller, and back-end T-SQL errors); deployment, configuration, and installation (code deployment, initial-data and server/tool installation, configuration, and other installation errors); and software maintenance (data-change-management errors and regression errors, where fixing one piece of code breaks another). A number of residual categories capture defects that fall outside the development team's direct control or are not genuine defects—test errors, end-user errors, defects originating in external systems, defects that could not be reproduced, and reports that were ultimately not bugs. This grouping serves two purposes: it keeps the fine-grained categories that make the analysis actionable, while allowing the distribution to be read at the coarser, phase level when that is more informative.

#### 3.4.3.2 AI-assisted classification, observations and insights

As in Analysis 1, the classification was carried out through human–AI collaboration under human oversight (Section 3.4.1). The categories themselves were defined by the author and the company's senior technical managers, drawing on established human-error and cognitive-performance frameworks—Rasmussen's skill–rule–knowledge model [15] and Reason's work on human error [16], and on human-error taxonomies specific to software defects [14], together with our own related application of these ideas in software engineering [17].

For records where the originating SDLC phase had been captured directly by the resolving engineer (Section 3.2), that judgement was used; for the remainder, the AI inferred the likely origin from the root-cause and fix descriptions, and every AI-inferred origin was reviewed and, where necessary, corrected by the author and the senior technical managers before being accepted. Because defect origin often depends on contextual and design knowledge, this human validation was essential rather than optional.

Figure 4 presents the distribution of defect origins for the 374 defects. We observed that a small number of defects were found to have more than one originating human error; each such defect is counted under each of its origins. As a result, the origin counts in Figure 4 sum to 389 rather than 374, the excess of 15 corresponding to the compound defects discussed above. The distribution should therefore be read as the distribution of defect origins, not of defects.

The dominant observation is that coding errors account for the large majority of origins—around 68%—and, within coding, that back-end errors substantially outweigh front-end ones: back-end controller logic (112) and back-end database/T-SQL code (67) together account for roughly twice as many origins as front-end code (86). This concentration is itself an actionable finding, and one that a conventional phase-level view would have hidden: treated as a single "implementation" phase, coding would appear only as the largest bucket, whereas the finer-grained scheme localizes the problem specifically to back-end logic and database code, pointing to where reviews, tests, and coding standards would have the greatest effect.

Beyond coding, the distribution reveals several smaller but instructive concentrations. Design-stage errors, though far fewer (26 in total, spanning requirements analysis, database design, and UI design), are consequential because errors introduced this early tend to surface much later and cost more to correct. Data-related origins are notable across the bands: database design errors (11), initial-data installation errors (13), and data-change-management errors during maintenance (19) together indicate that a meaningful share of defects stem not from application logic as such, but from decisions about data structure, seeding, and change—an origin that a purely code-focused analysis would miss entirely. The residual categories are analytically useful in their own right: the modest numbers of test errors (12), end-user errors (5), defects caused by external systems (12), non-reproducible reports (6), and non-defects (2) confirm that the large majority of records reflect genuine, internally-introduced defects, which lends confidence to the origin analysis as a basis for the improvements discussed in Section 3.5. At the same time, the presence of these categories illustrates a practical benefit of the classification: it distinguishes defects the development team can act upon from those originating with end users, external systems, or misreports, so that improvement effort is directed where it can have effect.

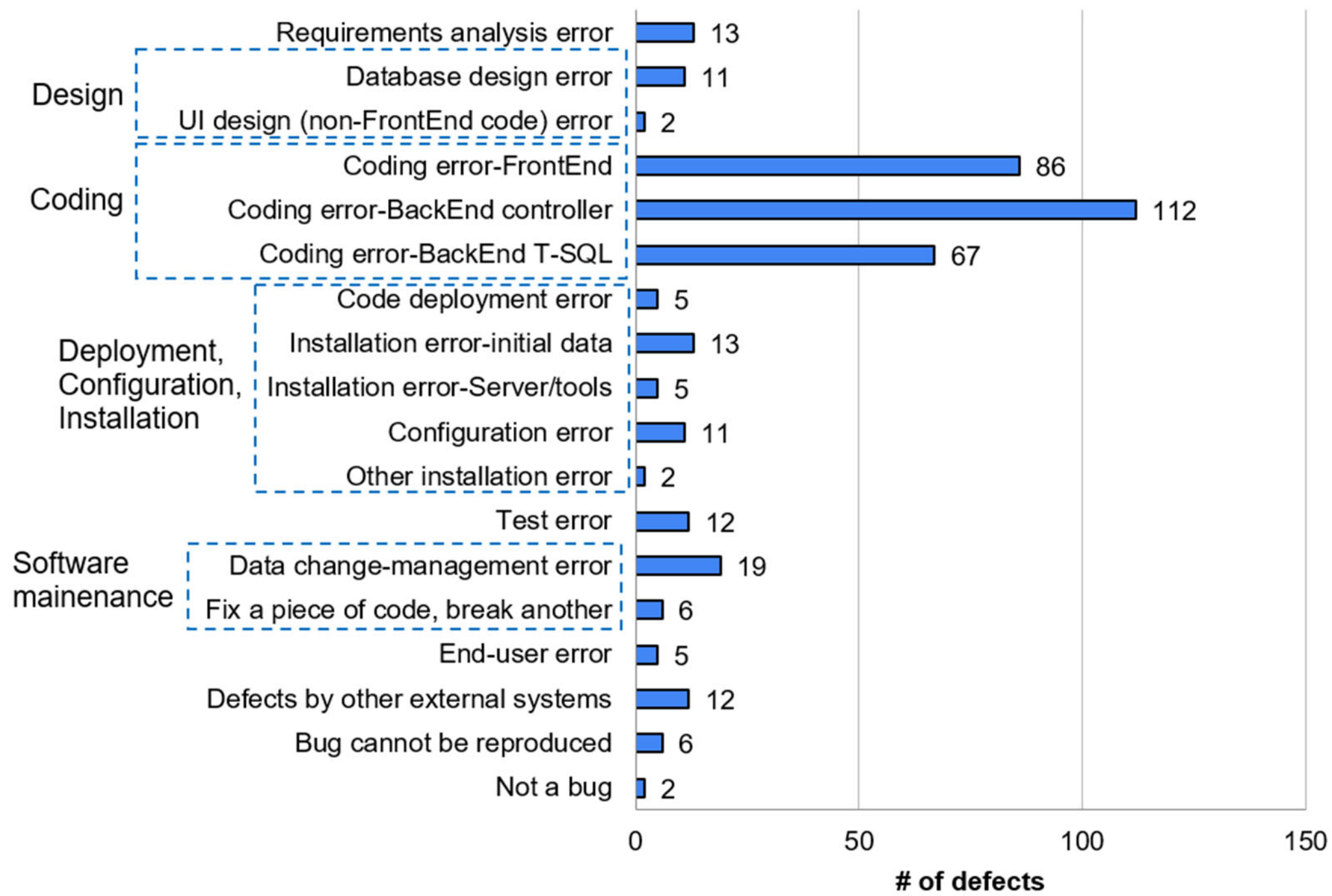

**Figure 4-Distribution of defect origins across SDLC phases: Types of Human errors leading to defects**

As discussed above, one noteworthy observation was that some defects originated from more than one human error. While most defects in the dataset could be traced to a single originating human error, we observed that a number of defects were caused by more than one distinct human errors acting together, frequently introduced at different points in development. In our records, such cases are represented by a defect that carries two source-phase / human-error classifications rather than one.

The cases we examined share a characteristic pattern: the multiple contributing human errors typically straddle a boundary between parts of the system or between development activities. In one case, a failure to save a document was traced to both a front-end fault (an incorrect parameter sent from the client) and a back-end fault (the save logic relying on an inappropriate identifier field). In another, a defect in a management module combined a database-design fault with a back-end coding fault, so that the fix required correcting both a database view and the corresponding data-retrieval methods. In a third, a set of incorrect payroll calculations stemmed from both a back-end stored-procedure fault and a data-change-management fault, in which a duplicated input caused values to be counted twice. Such compound defects are analytically important for two reasons. First, they complicate origin classification, since a single defect may legitimately be associated with more than one fault and one human error, and hence more than one SDLC phase. Second, they carry a clear prevention implication: a control targeting only one of the contributing faults may fail to prevent the defect, so defects of this kind are better addressed by controls that span the boundary—front-end/back-end, design/implementation, or code/data—at which the multiple faults combined. We return to their prevention in Analysis 3.

#### 3.4.4 Analysis #3 - Determining how each bug could have been prevented

Analysis 3 asks, for each defect, how it could have been prevented—what specific quality-assurance or development practice, applied at the right point, would plausibly have stopped it from being introduced or reaching production. It is the pivot between understanding defects and acting on them, and it feeds directly into the improvements in Section 3.5.

We conducted this analysis at a fine-grained, per-defect level: for each defect, the AI proposed a concrete preventive measure grounded in that defect's own root cause and fix, which was then reviewed by the author and the company's senior technical managers. Working per defect preserves each case's technical context and supports human oversight, since every suggestion can be checked against the specific defect it refers to. The resulting measures can then be aggregated along whichever dimension is useful—by defect type, by originating SDLC phase, or by system—to inform the improvements in Section 3.5.

Table 2 illustrates the analysis using the same four defects as Table 1, each extended with an AI-generated, human-reviewed preventive measure. The suggestions are specific and actionable rather than generic: pre-populating required status codes during initial-data setup (example bug #1), configuration review across environments (example bug #2), a deployment checklist verifying required server components (example bug #3), and form-state reset with front-end validation (example bug #4). More importantly, these four preventions are of different kinds—spanning initial-data setup, configuration management, deployment verification, and front-end validation—even though several of the underlying defects look similar at the failure level. This confirms a premise of the approach: because defects with comparable symptoms can have different origins, effective prevention is origin-specific and depends on the origin-level understanding from Analysis 2.

**Table 2-Several example records from the dataset: defect (bug reports) and defect-fix data (same bugs as the four examples in Table 1)**

| # | Bug (failure) description (from bug report) | Written after the defect was diagnosed and fixed | | |
|---|---|---|---|---|
| | | Root cause of the defect | How the defect was fixed | How could have this bug been prevented? (Generated by AI, reviewed by human) |
| 1 | Duplicating applications on the X page raises "Duplicate application status not found." | The "Duplicate" status row was missing from the related database status table (data not loaded/set up correctly). | Added the missing "DUPLICATE" row to the status table. | Pre-populate the database with all required status codes (including DUPLICATE) during initial setup and deployment validation. |
| 2 | Notification widget (top-right of screen) does not work. | Configuration error: notification_shown was set to 1 instead of 0; the code itself was correct. | Corrected the setting to notification_shown = 0. | Implement code review and configuration management to verify that critical settings (notification_shown=0) are correctly deployed in all environments. |
| 3 | After adding data, search fails with "Full-Text Search is not installed, or a full-text component cannot be loaded." | Incorrect MS-SQL installation: the Full-Text and Semantic Extractions for Search component was not installed. | Reinstalled MS-SQL with the Full-Text Search component added. | Include full SQL Server component validation in the deployment checklist—verify all required features (Full-Text Search) are installed before go-live. |
| 4 | Creating/copying a document in Funds–Advance Report creates two copies with the same numbering. | Code error: after saving, the form was not reset or closed, so the backend re-accepted the same data and created a duplicate. | Updated the code so that after saving, the user is redirected to the new document's edit screen. | Reset form state after Save action to prevent duplicate submissions; add front-end validation to block double saves within a time window. |

### 3.4.5 Analysis #4 - Extraction and synthesis of engineer-focused insights

The final analysis turns from defects themselves to the engineers whose work they relate to, with the aim of supporting individual learning and capability building (Section 3.5.3). It proceeds in two steps: first identifying which engineer or engineers were associated with each defect (Section 3.4.5.1), and then synthesizing, per engineer, the recurring patterns of errors in their work (Section 3.4.5.2).

#### 3.4.5.1 Putting proper safeguards in place for this analysis

Because this analysis involved individual-level data about the errors behind defects, it was the most sensitive part of the work, and it was conducted under a set of explicit safeguards. In summary, engineers were informed that defect and error data relating to their work would be analyzed for learning purposes and their own self-improvements, and the initiative was discussed openly within the teams; the data was used solely for reflection and targeted skill development, and never for performance appraisal, ranking, compensation, or disciplinary decisions. The analysis was carried out entirely on anonymized identifiers (of the form *DEVx*), with the identity key held separately and restricted to two senior managers. Errors were treated as signals about processes and capability gaps rather than as individual fault, consistent with a blame-aware view of human error [15-17]. Feedback derived from the analysis was provided to each engineer privately, concerning only their own work, with no cross-engineer comparison or ranking. These safeguards, together with their limitations, are described in full in Section 3.6.

#### 3.4.5.2 Finding out which engineer(s) committed a given error

Identifying which engineer resolved a given defect was straightforward: the issue-management system records the resolving engineer through a drop-down field, completed by whoever fixed the defect. Identifying which engineer committed the error that led to a defect, however, was considerably harder in most cases. In a complex source file where many lines are added or changed and then merged into the central repository, the data flow, control flow, and dependencies of a given method may give rise to a defect whose underlying error was introduced—and possibly propagated—across several changes made by several people.

To determine the originating engineer systematically, we again used AI assistance (Claude). We integrated the AI with the company's internal Git server and with the engineers' development environments (for example, VS Code), and instructed it to examine the codebase and related artifacts—change sets, file version histories, and their dependencies—to reconstruct the chain of events leading to the human error behind a reported defect. As with the other analyses, the AI's conclusions were treated as candidate findings subject to human review rather than as authoritative attributions.

Two complications, both following from the compound defects observed in Section 3.4.3, are worth noting. First, some defects originated from more than one human error, and in some cases those errors were made by different people—for instance, a fault in the back-end and a related fault in the front-end together producing a single reported defect; in such cases we recorded the identifiers of all engineers whose errors contributed. Second, resolving a single defect sometimes required fixes from more than one engineer—again, for example, a back-end and a front-end correction made by the two developers responsible for those respective modules. Table 3 illustrates both the resolution and origination attributions, using the same four defects as Tables 1 and 2.

To protect the individuals concerned, engineers are referred to throughout this paper—and in internal group discussions of the analysis—only by anonymized identifiers of the form $DEVx$. The key linking these identifiers to individuals is held solely by the company's senior managers, who alone can associate an identifier with a person; the analysis itself was conducted on the anonymized identifiers.

**Table 3-Several example defects (bug reports) and determination of engineers involved with them (same bugs as the four examples in Table 1)**

| # | Bug (failure) description (from bug report) | Written after the defect was diagnosed and fixed | |
|---|---|---|---|
| | | Engineer(s) who committed the error | Engineer(s) who resolved (fixed) the bug |
| 1 | Duplicating applications on the X page raises "Duplicate application status not found." | DEV12 | DEV1 |
| 2 | Notification widget (top-right of screen) does not work. | DEV15 | DEV8 |
| 3 | After adding data, search fails with "Full-Text Search is not installed, or a full-text component cannot be loaded." | DEV4 | DEV25 |
| 4 | Creating/copying a document in Funds–Advance Report creates two copies with the same numbering. | DEV3 | DEV25 |

#### 3.4.5.3 Determining the patterns of errors (mistakes) made by each engineer

Once defects are associated with anonymized engineer identifiers, the per-engineer error history can be synthesized to reveal recurring patterns. Rather than a single defect count, the aim is a qualitative profile of the kinds of errors that recur in an engineer's work, which can then inform targeted learning. For example, the synthesis might indicate that one engineer's defects cluster in back-end data-handling logic, particularly around unvalidated assumptions about database state; that another's recur in front-end input validation; or that a third's arise mainly during data-change-management activities in maintenance.

Such patterns are more useful for development than raw totals, because they point to a specific capability to strengthen—database-state reasoning, input validation, or change management, respectively—rather than simply indicating that an engineer was associated with a certain number of defects, many or few. How these per-engineer profiles were fed back to engineers, and the safeguards applied in doing so, are described in Sections 3.5.3 and 3.6.

## 3.5 Using the results of analytics for different improvements

The analyses described in Section 3.4 were valuable only insofar as they led to action. This section describes the improvements the analytics informed, along three streams—software process improvement, quality assurance, and engineer capability building—while the evaluation of their usefulness and benefits is reported in Section 4.

### 3.5.1 Suggested insights for Software Process Improvement (SPI)

The origin and prevention analyses (Sections 3.4.3–3.4.4) provided a basis for improving development processes by making visible where existing processes repeatedly allowed similar defects to arise. Rather than treating defects as isolated incidents, the organization used the aggregated origin patterns as a learning mechanism to locate systematic process weaknesses—that is, points at which the process itself, rather than any individual, permitted the same class of error to recur. This reframing is important: a single defect invites a single fix, whereas a recurring origin invites a change to the way work is done.

Several concrete process insights emerged. The concentration of defects originating in back-end and database logic, together with the recurring appearance of data-related origins spanning database design, initial-data installation, and data-change management (Section 3.4.3), pointed to a process gap in how data-related design decisions were discussed and documented before implementation. In response, the organization refined its practices so that data assumptions and data-related design decisions were made explicit and reviewed earlier, before coding began. A second insight concerned the design stage: although design-origin defects were comparatively few in number, several of them—particularly requirements-analysis and database-design errors—were associated with defects that surfaced much later and required disproportionately large fixes, which motivated earlier and more structured review of requirements and data models. A third insight came from the maintenance band: the recurrence of data-change-management errors and of regressions in which fixing one defect introduced another (Section 3.4.3) indicated a need for more disciplined handling and verification of changes to production data and code. In each case, the value of the analytics was not that any single insight was individually surprising, but that the aggregated, origin-level evidence allowed process-improvement effort to be prioritized against the origins that actually recurred, rather than against assumptions about where problems lay. The usefulness of these process changes is evaluated in Section 4.

### 3.5.2 Shift-Left QA: Moving from defect detection to defect prevention practices

The same origin-level understanding was used to strengthen and re-target quality assurance, shifting the emphasis from late-stage detection toward earlier and more focused controls—a "shift-left" of assurance effort [11]. The guiding principle was to place controls at, or close to, the point where the human errors behind defects were being introduced, so that defects would be prevented or caught earlier rather than discovered in testing or production. Because the analytics localized defect origins finely—for example, to back-end controller logic and to database code, rather than to "implementation" in general—assurance measures could be directed specifically rather than applied uniformly across all development.

This produced a set of targeted assurance changes, each tied to a recurring origin. Recurring back-end and database origins motivated stronger and more focused code review and static analysis, concentrating reviewer and tool attention on conditional logic, data handling, and database code, where the analytics showed errors were most often introduced. Recurring requirements- and design-stage origins motivated earlier review activities, including structured review of requirements and validation of data models and interface designs before implementation. Recurring configuration, installation, and deployment origins motivated tighter deployment verification, so that environment- and data-setup errors—of the kind illustrated in Table 1—would be caught before release rather than surfacing in the live system. The common thread was selectivity: instead of expanding testing uniformly, the organization moved assurance earlier and concentrated it where the origin analysis indicated it would have the greatest effect. We deliberately make no claim here about resulting changes in defect volume.

### 3.5.3 Capability building and upskilling through periodic feedback

The engineer-focused synthesis (Section 3.4.5) was used to support individual learning through periodic, private feedback. Drawing on the per-engineer patterns of recurring error types, the analytics pipeline generated individual summaries highlighting the kinds of issues that recurred in an engineer's own work—for example, a tendency toward defects in back-end data handling, in front-end input validation, or in data-change management—together with illustrative examples and, where relevant, the corresponding preventive practices identified in Analysis 3 (Section 3.4.4). The intent was developmental: to help each engineer notice a recurring pattern in their own work and focus attention on a specific capability to strengthen. Feedback was delivered privately to the individual concerned, described only that engineer's own defects, and contained no comparison to colleagues, no ranking, and no aggregate score, consistent with the safeguards in Sections 3.4.5 and 3.6. For the purpose of this pilot project, the bug dataset and systems under analysis, 28 developers were involved.

In practice, the feedback took the form of a short periodic message summarizing the engineer's most frequent recurring error area, one or two concrete examples drawn from their own resolved defects, and a specific, forward-looking suggestion for improvement. Two illustrative messages (anonymized, with identifiers of the form DEVx) convey the style:

> A feedback email sent to DEV7: *Over the recent period, the most common recurring pattern in defects associated with your work relates to back-end data handling—specifically, assumptions about the state of the database that did not always hold at run time (for example, expecting a required status row or reference record to be present). It may be worth giving particular attention, during coding and review, to validating these data assumptions explicitly rather than relying on them.*

> A feedback email sent to DEV12: *Over the recent period, a recurring pattern in defects associated with your work relates to front-end input validation—cases where input was accepted or rejected incorrectly, or where a validation response was missing.*

*This is a well-defined and improvable area; focusing on consistent input-validation handling on the client side, and reviewing the example cases below, may help reduce recurrence. Please treat this as a prompt for reflection rather than an assessment.*

Feedback of this kind was intended to make an otherwise invisible pattern visible to the individual, so that learning could be self-directed. Because the messages drew only on the engineer's own history and were expressed developmentally, they were designed to support reflection and skill-building rather than evaluation. The perceived usefulness of this feedback to engineers—including whether they found it accurate, actionable, and appropriate—is evaluated in Section 4.

### 3.6 Ethics, confidentiality, and responsible use of developer-level data

Because part of this work involved analyzing data about the errors behind defects at the level of individual engineers (Section 3.4.5), it raised ethical considerations that we addressed deliberately rather than incidentally. We describe here the safeguards applied, and we are explicit about their limits.

The overriding principle was that the analysis served learning, not evaluation. Engineer-level error data was used solely to support reflection and targeted skill development, and was never used for performance appraisal, ranking, promotion, compensation, or disciplinary decisions. This purpose limitation was not merely stated but built into how the data was handled and how feedback was framed, as described below.

- *Awareness:* Engineers were informed that defect and defect-fix data relating to their work would be analyzed for learning and process-improvement purposes, and the initiative was discussed openly within the teams rather than conducted covertly.
- *Anonymization and access control:* Throughout the analysis, and in all internal discussions of it, engineers were referred to only by anonymized identifiers of the form DEVx. The analysis itself was conducted entirely on these identifiers. The key linking identifiers to individuals was held separately and restricted to a small number of senior managers.
- *No comparison or ranking.* Feedback derived from the analysis was individual-only: each engineer received information concerning solely their own recurring error patterns, with no comparison to colleagues, no team-relative figures, and no aggregate scores.
- *Developmental framing and right of response:* Feedback was delivered privately and framed as a prompt for reflection, and engineers were able to discuss or contest it. Errors were treated as signals about processes and capability gaps rather than as individual fault, consistent with a blame-aware view of human error in which the aim is to improve the system of work rather than to attribute blame [15-17].
- *Data governance.* The AI-assisted analyses used commercial, cloud-based large language model services (Section 3.4.1.1); the textual data required for each analysis was therefore transmitted to these external services. Several measures governed this. Individuals were referred to only by anonymized identifiers of the form DEVx, and client, system, and project names were replaced with placeholders before data was submitted. The data submitted for the classification and prevention analyses consisted principally of the textual defect fields—defect descriptions, root-cause explanations, and fix descriptions—in anonymized form. The engineer-attribution step (Analysis 4) used the company's existing integration of Claude Code with its Git repositories, which is part of the development team's standard workflow; and under that workflow, engineers are instructed to anonymize or omit sensitive portions of code when working with the tool. Throughout, the services were used under a commercial license whose terms provide that submitted data is not used to train the providers' models. There are plans in the company to acquire an on-premise or Zero Data Retention (ZDR) AI platform. Organizations with stricter data-governance requirements may require such deployments to adopt the approach.

Two limitations of these safeguards should be stated plainly. First, the anonymization is reversible by design: because the identifier key is held by senior managers, those managers can in principle associate an anonymized profile with a named individual. The protection engineers have against the use of this data for evaluation therefore rests not only on anonymization but also on the organizational commitment to the purpose limitation described above; anonymization alone does not guarantee it. Second, the same senior managers participated in defining and validating the classifications (Sections 3.4.2–3.4.3), which concentrates both analytical authority and the ability to de-anonymize in a small group. We judged this acceptable in an industrial action-research setting where domain and system knowledge were essential to the analysis, but we acknowledge it as a constraint on the independence of the individual-level analysis, and we return to its implications for validity in Section 5. We report these limitations openly because, in our view, responsible use of developer-level data depends as much on acknowledging what safeguards cannot guarantee as on the safeguards themselves.

## 4 EVALUATING THE USEFULNESS AND BENEFITS OF THE APPROACH

Having described the approach and how its results were used (Section 3), we now evaluate it. This section sets out the case-study design (Section 4.1), the research questions and how they were answered (Section 4.2), the dataset and participants (Sections 4.3–4.4), and the analysis procedure (Section 4.5); the results follow in Section 4.6, and the approach's adoption and current use in the company in Section 4.7.

### 4.1 Case-study design and its relation to the overall research approach

We evaluated the approach through an industrial case study conducted in the organization where it had been deployed, following established guidelines for case-study research in software engineering [28]. The study is of the *improving* type: the approach was introduced into live development settings, and the evaluation examines both how it performed and what changed as a result. This is consistent with the action-research and technology-transfer setting described in Section 2.1, in which the researcher participated in introducing the approach rather than observing it from outside—a position that offers close access and contextual insight but also carries known risks to objectivity, which we address through triangulation below and revisit in Section 5.

We deliberately did not adopt a controlled experimental design. Defect volumes in the studied organization are shaped by feature scope, release cadence, team composition, and system maturity, none of which could be held constant, and no comparable control setting was available. We therefore designed a triangulated evaluation drawing on three largely independent sources of evidence—practitioner judgement (via survey and interviews), the classification-reliability checks, and the defect corpus itself together with documented process and tooling changes, and the defect corpus itself together with documented process and tooling changes—so that no central conclusion rests solely on the assessment of the researcher who introduced the approach.

### 4.2 Goal, Research questions, and Metrics

We structured the evaluation using the Goal–Question–Metric (GQM) paradigm [29]: a single measurement goal, refined into three research questions, each associated with concrete metrics and data sources. We use GQM here as an organizing structure rather than as a heavyweight measurement program, to keep the evaluation focused and to make explicit how each question is answered.

Following the standard GQM goal template, the measurement goal of this study is:

> <u>To</u> *analyze the AI-assisted defect and defect-fix analytics approach*
> <u>*for the purpose of*</u> *evaluation,*
> <u>*with respect to*</u> *its accuracy, usefulness, and prevention potential,*
> <u>*from the point of view of*</u> *the software engineers and technical managers who use it,*
> <u>*in the context of*</u> *a large software company developing web-based systems for government, finance, and healthcare.*

This goal is refined into three research questions:

- RQ1 — How accurate and reliable are the AI-assisted classifications and analytics outputs?
- RQ2 — How useful and actionable do practitioners find the analytics and the improvements derived from it?
- RQ3 — Is there quantitative evidence that engineers' defect density changed following the introduction of periodic feedback, in particular for the error types the feedback targeted?

Table 4 provides the GQM structure by associating each question with its metrics, data sources, and analysis methods.

**Table 4-GQM structure: research questions, metrics, and data collection**

| RQ | Metric(s) | Data source / instrument | Analysis |
|---|---|---|---|
| RQ1 | Proportion of original classifications confirmed correct on joint team review; percentage agreement and Cohen's κ between an independent blind re-classification and the original assignments; practitioner-rated accuracy of classifications for their own defects | Joint review of a stratified sample (71 defects); independent blind re-classification of a smaller sub-sample (55 of the 71 defects) by an uninvolved senior manager; survey items | Confirmation rate; percentage agreement; Cohen's κ, and 95% confidence intervals; descriptive summary |
| RQ2 | Practitioner-rated usefulness and actionability of the analytics outputs and of the process, QA, and capability-feedback improvements; reported changes in practice | Anonymous survey (Likert + free-text); semi-structured interviews | Descriptive statistics; thematic analysis |
| RQ3 | Per-developer defect density (defects per task completed and defects per actual hour worked), pre versus post the introduction of periodic feedback, overall and for feedback-targeted error types | Defect records with commit dates; company task and timesheet (actual-hours) records | Pre/post comparison of normalized defect density; descriptive, reported as indicative |

The instruments underlying Table 4 are described in Section 4.4 and the analysis procedures in Section 4.5. Evidence of adoption and sustained use, drawn from internal records and confirmed with the senior technical managers, is reported separately in Section 4.7.

Two points about scope should be stated at the outset. First, the evaluation is a single-organization industrial case study without a control group; RQ3 in particular is a single-group, pre/post (quasi-experimental) comparison, and we therefore report its results as evidence *consistent with* impact rather than as proof of a causal effect attributable to any single element of the approach. Second, because the periodic feedback was introduced alongside the process and quality-assurance changes of Section 3.5, any improvement observed in RQ3 reflects the approach as a whole rather than the feedback in isolation. We return to these limitations in Section 5.

### 4.3 The defect dataset under study

The findings and insights reported in this paper are based on a subset of 374 defect records, drawn from a substantially larger body of defect data accumulated in the organization's issue-tracking system over several years (Section 3.2). Because the construction of this dataset affects how its results should be interpreted, we describe the selection procedure, inclusion and exclusion criteria, handling of missing data, and composition in detail.

*Selection procedure and inclusion/exclusion criteria.* The dataset was assembled from defect records that met the following criteria: (i) the defect record should correspond to a resolved (fixed) defect, so that both the defect and its fix were available for analysis; (ii) the defect record should be related to one of the six selected production software systems within the scope of the study. We shall clarify that these systems were chosen since they were actively maintained by the participating teams during the study period; and (iii) the defect record should contain the textual root-cause and resolution fields on which the analyses depend. Records were excluded if they were unresolved or still-open at the time of data extraction; were duplicates of another record; or were lacking any root-cause/fix description. The chosen dataset (374 defect records) was a random sample drawn from the larger set of defect records that met the above criteria within the chosen time window (2024–2026). The sampling mechanism was as follows: records were selected using a random number generator over the eligible record IDs. Because the sample was drawn at random from all eligible records, it is expected to be representative of the larger population of resolved defects in the studied systems over this period, without selection bias among the eligible records toward particular systems, defect types, or origins. But it should be noted that the eligible population is itself restricted to six selected systems and to records containing the required root-cause/fix information. Data about developers, tasks, and hours used in RQ3 were taken from the company's internal task management system.

*Handling of missing and incomplete data.* As discussed in the inclusion/exclusion criteria above, records missing a root-cause or fix description were excluded rather than imputed.

*Distribution across systems and years.* Table 5 shows the composition of the dataset across the six systems (named S1 to S6, anonymously) and across the calendar years from which records were drawn (2024–2026). As we can see, majority of the records fall in 2025, reflecting the period of most active phase of the study.

**Table 5-Composition of the 374-defect dataset across systems and years**

| System | 2024 | 2025 | 2026 | Total |
|---|---|---|---|---|
| S1 | 6 | 20 | 8 | 34 |
| S2 | 15 | 44 | 18 | 77 |
| S3 | 9 | 30 | 12 | 51 |
| S4 | 13 | 38 | 16 | 67 |
| S5 | 18 | 55 | 23 | 96 |
| S6 | 8 | 29 | 12 | 49 |
| **Total** | **69** | **216** | **89** | **374** |

*Proportion classified manually versus by AI.* The two classifications differ in how they were produced, which is relevant to interpreting them. For the failure-type classification (Analysis 1), all 374 records were classified with AI assistance and then reviewed by the research team (Section 3.4.2); there was no manual developer-provided failure type. For the origin classification (Analysis 2), the provenance was mixed. For 253 of the 374 records (68%), the originating SDLC phase /

human-error type had been entered manually by the resolving engineer through the drop-down field introduced during the project (Section 3.2). For the remaining 121 records (32%), which predated the field, the origin was inferred by AI from the root-cause and fix descriptions and then validated by the researcher against the record; all of these AI-inferred origins were judged correct on validation. Note that this is distinct from the independent reliability assessment of the origin classification as a whole, reported in Section 4.6.1, which was conducted on a separate sample.

## 4.4 Participants and study instruments

The evaluation drew on the engineers and technical staff of the studied organization who were involved in, or affected by, the approach: software developers whose work was represented in the defect dataset, together with senior technical managers responsible for quality and development. In total, 28 developers and two senior technical managers were within scope of the study.

Three instruments were used, complemented by the company's existing task and timesheet records. The anonymous practitioner survey combined Likert-scale items—covering the perceived accuracy of the failure and origin classifications for respondents' own defects, and the usefulness and actionability of the analytics outputs and the derived improvements—with free-text items inviting elaboration and criticism.

The survey was completed by 19 of 28 eligible engineers (approximately 68% response rate). The semi-structured interviews, lasting approximately 45 minutes, were conducted with seven participants spanning developers, quality-assurance staff, and senior technical managers, and explored why particular insights and feedback were or were not useful, and what changed in practice as a result; interviews were recorded and transcribed with consent. The structured review exercise was conducted over a stratified sample of 71 of the 374 defects, drawn to span the range of systems and origin types, and is described in Section 4.5. The full practitioner survey instrument and semi-structured interview guide are provided in Appendices A and B.

For RQ3, the evaluation additionally used the company's task and timesheet records. The organization operates a mature effort-estimation and time-logging practice, refined over more than a decade: for each task, an estimated duration is agreed between the developer and manager beforehand, and the actual hours worked are logged on completion. Logged hours are routinely reviewed by managers for consistency, which maintains the discipline and reliability of the effort data independently of, and long prior to, this study. These records allow per-developer output to be quantified by both the number of tasks completed and the actual hours worked, providing the denominators used to normalize defect density in RQ3.

*Ethics and anonymity.* Participation in the survey and interviews was voluntary, and respondents could decline or withdraw without consequence. The survey was anonymous, and interviews were recorded and transcribed only with the participant's consent. No personally identifying information is reported; all engineer-level data, including the task and timesheet records used for RQ3, were handled under the anonymization and confidentiality safeguards described in Section 3.6, and the individuals, systems, and organization are referred to only in anonymized or generalized form throughout.

## 4.5 Analysis procedure

The analysis combined quantitative and qualitative procedures, kept deliberately simple given the sample sizes involved.

*Classification accuracy and reliability (RQ1).* Two complementary checks were applied over the stratified sample of 71 defects. First, the author and the senior technical managers jointly re-examined each sampled defect: the original classification—the AI-generated failure type (Analysis 1) and the origin/human-error type, which for some records had been assigned by the resolving engineer and for others inferred by the AI (Analysis 2)—was reviewed against the defect's description, root cause, and fix, and judged either correct or requiring correction. We report the confirmation rate: the proportion of original classifications confirmed correct on review. Because this joint review was performed by members of the research team and was not blind to the original labels, a second, independent check was added: one senior technical manager who had not been involved in developing the classification schemes independently re-classified a smaller blind sub-sample of 55 of the 71 defects, without sight of the original labels, and agreement with the original assignments is reported as a percentage agreement. Because the joint review was not blind, we regard the independent blind re-classification as the primary reliability estimate, with the joint review reported as a measure of the quality of the final (corrected) classifications rather than of raw AI output.

*Perceived usefulness (RQ2).* Survey Likert responses were summarized as distributions. Interview transcripts and survey free-text responses were analyzed thematically, with codes grouped into themes that were checked against the underlying source material.

*Impact on subsequent work (RQ3).* To examine whether the approach was associated with improvement in engineers' work, we compared defect density before and after the introduction of periodic feedback. For each developer, we defined two three-month windows—the three months before and the three months after feedback began—and computed defect density in each. A three-month window was chosen deliberately: shorter periods (such as a single month) contain too little coding and testing activity per developer to yield a stable measure, whereas three months allow enough work to accumulate and be tested. To avoid the distortion that defect recording dates would introduce, defects were assigned to windows using the commit date of the originating code rather than the date a defect was recorded.

Because a raw defect count reflects how much a developer produced rather than the quality of their work, we normalized it using the company's task and timesheet records (Section 4.4), expressing defect density in two independent ways: defects per task completed, and defects per actual hour worked. Reporting both provides a robustness check—a change that appears consistently under both denominators is more credible than one visible under only one. Density was computed both overall and, separately, for the specific error types that each developer's feedback had targeted, since an improvement concentrated in the flagged error types is more indicative of a feedback-related effect than a general change in defects. The analysis included the developers who were sufficiently active in both windows for the density measure to be meaningful—19 of the 28 developers; the remainder were excluded mainly because they had joined recently, worked primarily on systems outside the studied set, or were absent for part of the period. We report the resulting changes descriptively and interpret them as indicative rather than conclusive, for the reasons given in Section 4.2.

*Use of AI in the analysis:* Consistent with the approach itself, generative AI assisted the thematic coding as an analytical aid only; all AI-produced codes, themes, and summaries were inspected and validated by the author against the source material, and no finding was accepted that could not be traced to specific underlying data.

## 4.6 Results

### 4.6.1 RQ1: Accuracy and reliability of the classifications and analytics outputs

RQ1 concerns whether the classifications and analytics outputs can be trusted. Across the stratified sample of 71 defects, the joint review by the author and the senior technical managers confirmed 91% of the original failure-type classifications and 88% of the origin classifications as correct, with the remaining cases requiring correction—most often the reassignment of a defect between two adjacent origin categories (for example, between back-end coding and database-related origins).

Because this joint review was carried out partly by those who produced the classifications, it measures agreement with the final, corrected labels rather than the accuracy of the original AI output; we therefore treat the independent blind re-classification below, not the joint review, as the primary evidence of reliability.

On the smaller blind sub-sample of 55 defects re-classified independently by an uninvolved senior manager, agreement with the original assignments was 85% for failure types (Cohen's $\kappa = 0.84$) and 80% for origin types ($\kappa = 0.76$), corresponding to "almost perfect" and "substantial" chance-corrected agreement, respectively [30]. The 95% confidence intervals were [74%, 92%] and [68%, 88%].

The few disagreements were concentrated at the boundaries between adjacent categories—most often between Data Persistence and Data Retrieval among failure types, and between back-end coding and database-related origins—rather than being spread across unrelated categories.

These figures are corroborated by practitioners' own judgements. In the survey, 84% of respondents agreed or strongly agreed that the failure-type classifications of their own defects were accurate, and 79% that the identified origins were accurate. Figure 5 shows the three accuracy signals side by side for failure and origin classifications. In summary, the internal review, the independent blind check, and the practitioner ratings indicate that the classifications were accurate enough to serve as a basis for the improvements in Section 3.5, while the disagreements—concentrated at the boundaries between related categories—mark the limits of that accuracy and are revisited in Section 5.

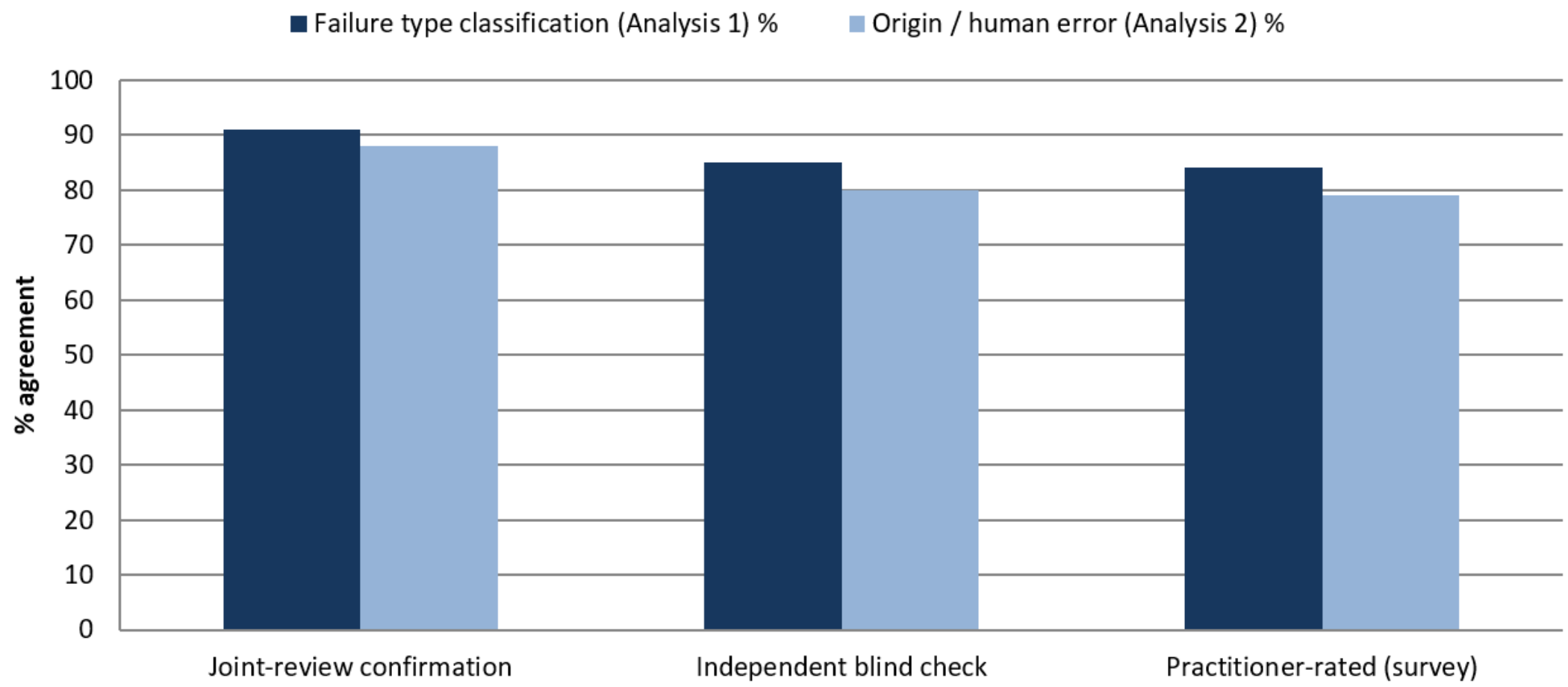

**Figure 5-Accuracy and reliability of AI-made classifications (RQ1)**

#### 4.6.2 RQ2: Perceived usefulness and actionability of results

RQ2 concerns whether practitioners found the analytics and derived improvements useful and actionable. Survey responses were positive across the three improvement streams. Figure 6 shows the distribution of responses for each usefulness item.

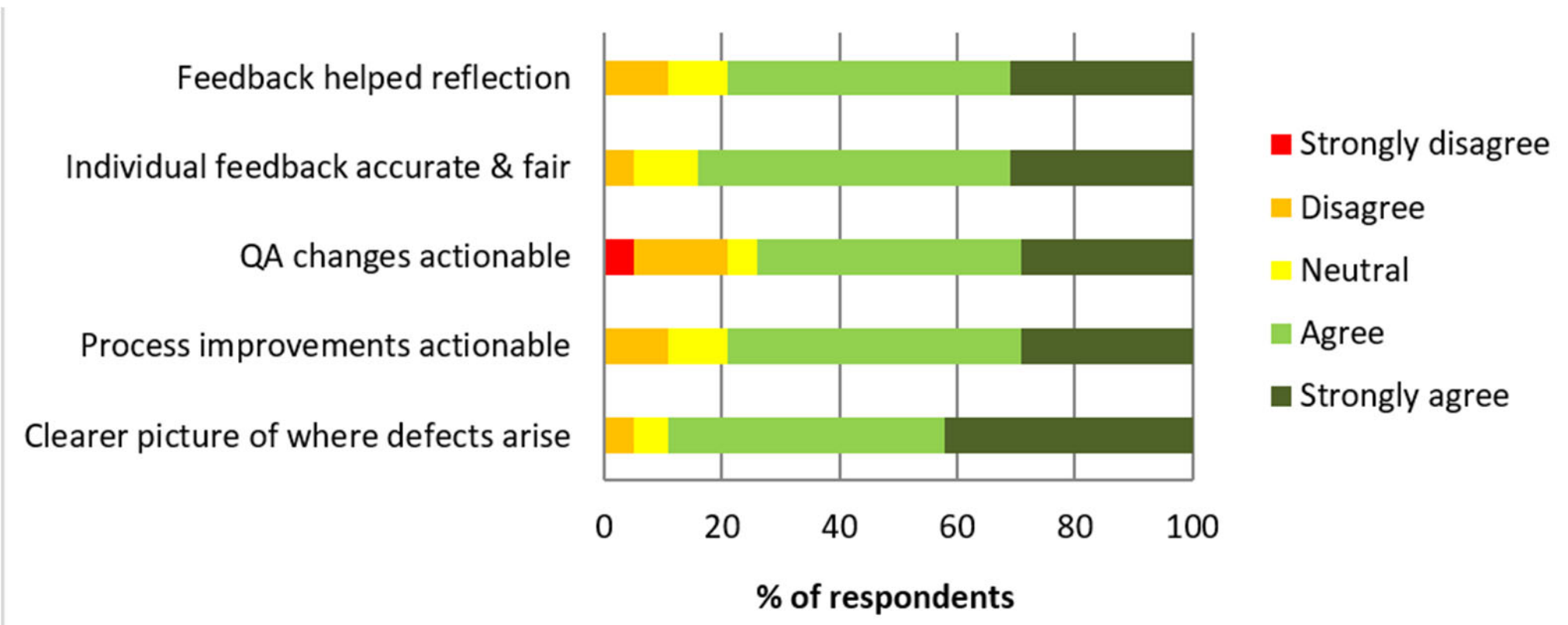

**Figure 6-Perceived usefulness and actionability of results (RQ2)**

Combining "agree" and "strongly agree," 89% of respondents felt the analytics gave a clearer picture of where and how defects arise than the team had before; 79% found the suggested process improvements relevant and actionable; 74% found the quality-assurance changes relevant and actionable; and 84% found the individual periodic feedback accurate and fair, with 79% agreeing it helped them reflect on their work. The lowest agreement was on the QA changes, where 21% disagreed or were neutral—suggesting these were seen as worthwhile but harder to act on without additional tooling and time.

The interviews explained these ratings. Several themes recurred. Participants valued the shift from isolated defects to visible patterns: "*Before, every bug was its own little fire. Seeing them grouped by where they came from changed how we talked about them*" (Developer). The origin-level view was seen as actionable precisely because it was specific: "*Knowing it was the back-end data handling, not just 'a bug', told us where to put the review effor*t" (Senior technical manager). On the individual feedback, the developmental framing mattered: "*It didn't feel like being marked—it was more like, here's a pattern in your own work, worth a look*" (Developer). Not all feedback was positive; some noted limits: "*A few of the suggestions were things we already knew; the value was in the ones we didn't*" (QA engineer). One manager placed the value at the team level: "*The biggest change wasn't any single fix—it was that we started aiming our reviews instead of spreading them thin*" (Senior technical manager).

#### 4.6.3 RQ3: Impact of the Feedbacks provided by approach on subsequent defect density of work done by developers

RQ3 asks whether there is quantitative evidence that engineers' work improved following the introduction of periodic feedback. As set out in Section 4.5, we compared normalized defect density in the three months before and after feedback began, for the 19 developers active in both windows, using two independent denominators and examining both overall defects and the specific error types each developer's feedback had targeted.

A word on the choice of denominator is warranted, since defect density is most commonly expressed per thousand lines of code (KLOC) [31]. We deliberately did not use a KLOC-based defect density measure. Much of the work in the studied context involves modifying, refactoring, or removing existing code rather than writing new code, which makes lines of code an ambiguous and potentially misleading measure of how much work a developer actually performed: a change touching few lines may represent substantial effort, and a large mechanical edit little. Attributing defects to added, changed, and deleted lines separately would further complicate the measure without clarifying it. We therefore normalized instead by two quantities that were reliably available and that more directly reflect the work undertaken—the number of tasks completed and the actual hours worked (Section 4.4). It should also be noted that the defects counted here are *reported* defects that reached testing or later stages, not all defects introduced and caught during development; the resulting densities are accordingly lower than the total-defect figures often cited in KLOC-based benchmarks, and are intended for within-team, pre/post comparison rather than cross-organization benchmarking, since task and effort calibration is specific to this organization.

At the group level, mean defect density decreased under both denominators from the pre- to the post-feedback window: from 0.42 to 0.34 defects per task, and from 0.048 to 0.040 defects per actual hour (Figure 7). The decrease was more pronounced for the feedback-targeted error types than for defects overall—targeted-type density fell from 0.22 to 0.14 defects per task, a proportionally larger reduction than the overall figure. That the change appears consistently under both denominators, and more strongly in the targeted error types, is what makes it more than a by-product of how much each developer happened to produce.

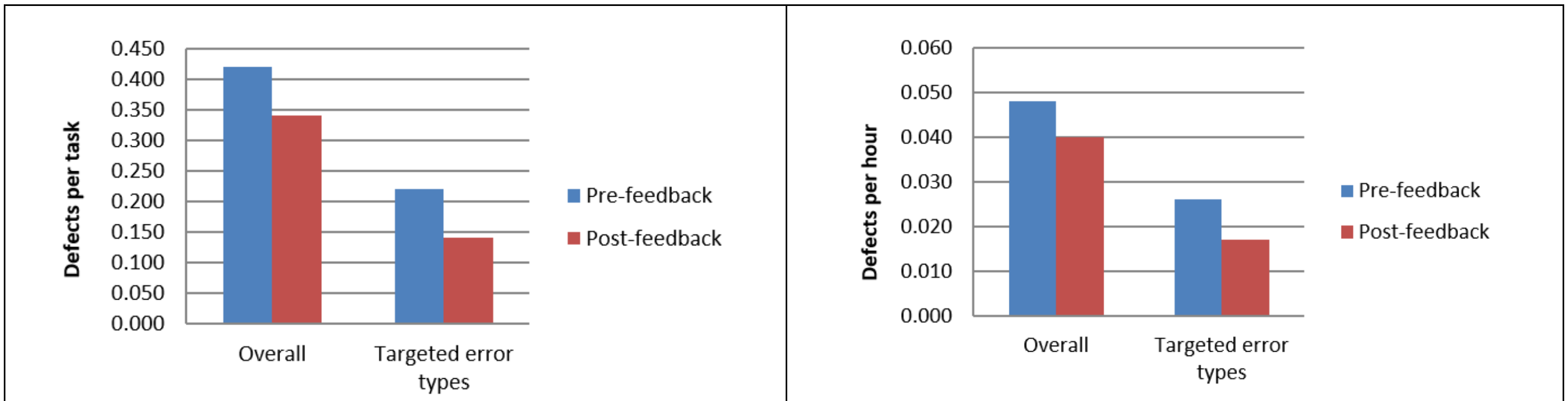

Defect density per task

Defect density per actual hour

**Figure 7- Impact of the feedbacks provided by the approach on subsequent defect density of work done by developers (RQ3)**

At the developer level, Figure 8 shows the change in defect density for each of the 19 developers individually, rather than for selected cases. Defect density decreased for 16 of the 19 developers following the introduction of feedback and increased for the remaining three, labeled as *dev_a, dev_b* and *dev_c* in Figure 8.

For those three developers, the increase coincided with a rise in the complexity of some of their assigned work over the study period (the researcher did make that observations by checking their work-log data), and possibly a lower engagement with the provided improvement feedback in the previous cycle; a possibility that we were not able to check with those developers, given the nature of the issue. The group-level decrease reported above therefore reflects a broad pattern across most developers rather than a few large individual improvements, while the presence of developers who did not improve underscores that per-developer density is shaped by task difficulty and individual factors as well as by the feedback.

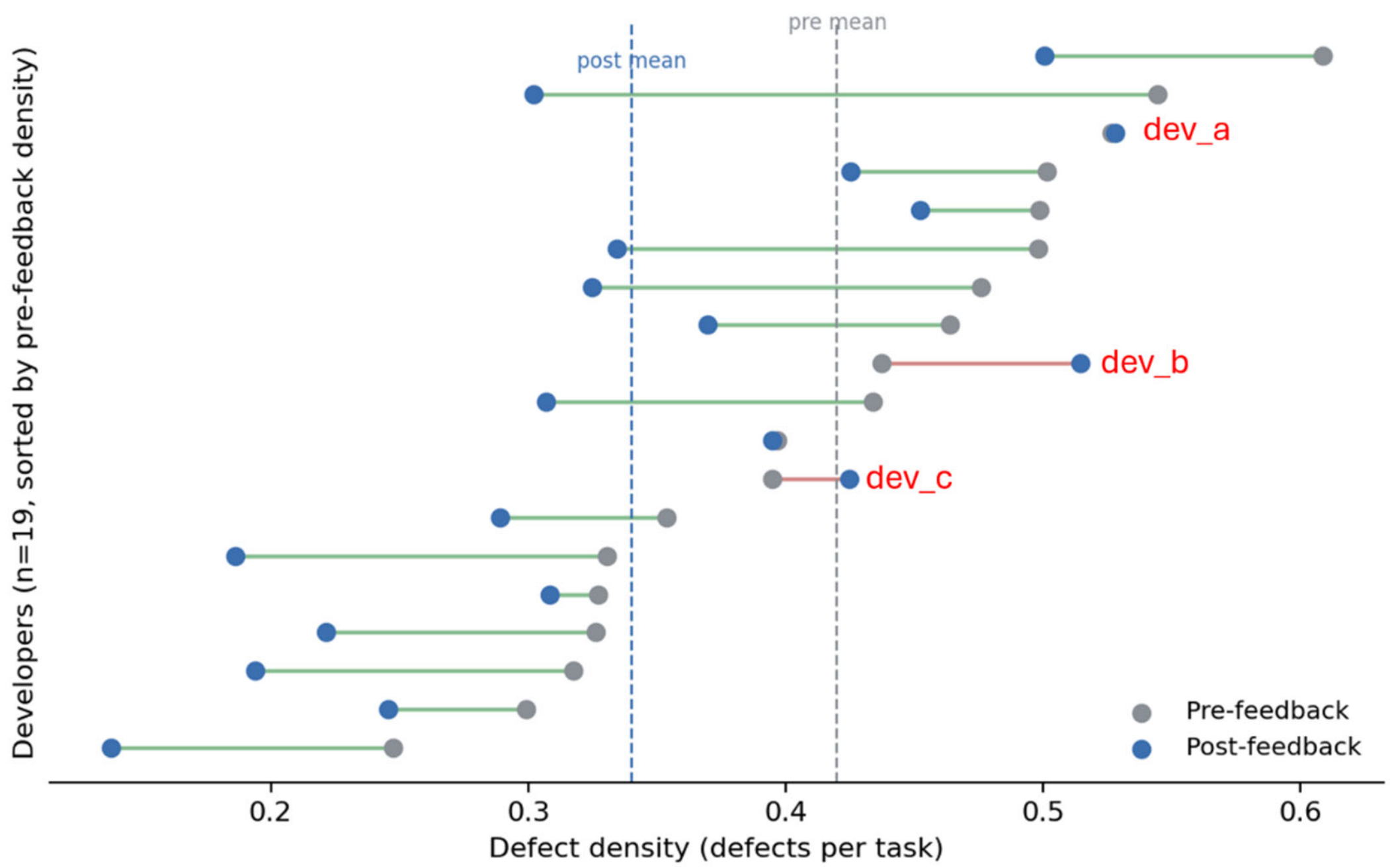

**Figure 8-A dumbbell plot showing per-developer defect density before and after the introduction of periodic feedback, for all 19 developers in the analysis. Each row is one developer; the grey and blue dots mark that developer's pre- and post-feedback defect density (defects per task), connected by a line colored green where density decreased and red where it increased. Density decreased for 16 of the 19 developers and increased for three (*dev_a, b, c*); only very slightly for one developer: *dev_a*. Dashed vertical lines mark the group means before (0.42) and after (0.34). Developers are sorted by pre-feedback density.**

To make this concrete at the record level, Table 6 shows two representative developers, DEVx and DEVy—chosen for their typical activity levels and regular receipt of monthly feedback, not for the size of their improvement—across the six-month period. For DEVx, the count of the flagged error type fell across successive months while overall task output was maintained; DEVy shows a similar but weaker pattern. These cases illustrate the group-level trend rather than establishing it; the evidence for RQ3 rests on the aggregate comparison across the 19 developers.

The observed change should be read with the design in mind: as a single-group, pre/post comparison without a control, introduced alongside the process and QA changes of Section 3.5, it cannot be attributed to the feedback alone or separated from concurrent factors such as growing developer experience, increasing maturity of the systems themselves, changes in workload or testing intensity, and the parallel process and QA changes..

The interviews support a cautious reading: asked what they attributed any improvement to, participants pointed to a combination of the feedback and the strengthened, better-targeted reviews rather than to the feedback in isolation. We therefore interpret this result of RQ3 as a tentative temporal association between the introduction of the approach and a subsequent reduction in defect density, rather than as strong evidence of a causal effect; the uncontrolled, single-group design cannot rule out maturation, workload change, system maturity, or the concurrent interventions as alternative explanations. The per-developer distribution in Figure 8 shows this reduction was broad rather than driven by a few individuals, while also showing the developers for whom density did not fall. Given the small number of developers (19) and the modest defect counts per developer, we judged per-developer confidence intervals and paired effect sizes to be insufficiently stable to report reliably; we therefore present the per-developer distribution in Figure 8 and adopt the tentative framing above to characterize the strength, and the limits, of this evidence.

Beyond the measured change, participants noted a qualitative benefit: the feedback fostered accountability and self-directed improvement, prompting some engineers to pursue targeted training, self-study, or certification in the areas their feedback had highlighted: "*After seeing the pattern, I took a short course on the exact thing I kept getting wrong*" (Developer). This suggests the approach's value extended beyond the defects it measured into how engineers directed their own development, though we report it as reported experience rather than a measured outcome.

**Table 6- Monthly defect attribution for two representative developers over the six-month period**

| Developer | Time window | # of tasks completed | Actual hours | Total defects | Defects committed in flagged error type | Defects / task | Defects / hour |
|---|---|---|---|---|---|---|---|
| DEVx | Months 1–3 (pre) | 24 | 205 | 10 | 6 | 0.42 | 0.049 |
| | Months 4–6 (post) | 26 | 220 | 8 | 3 | 0.31 | 0.036 |
| DEVy | Months 1–3 (pre) | 19 | 168 | 8 | 4 | 0.42 | 0.048 |
| | Months 4–6 (post) | 21 | 176 | 7 | 3 | 0.33 | 0.040 |

### 4.7 Deployment and Current Use in the Company

Beyond the evaluation results, it is worth describing how the approach fits into the organization's development practice and its current status, since sustained real-world use is itself a signal of value.

The organization follows an Agile development approach, working in two-week sprints. Its day-to-day process is mature and disciplined; while the company has not undergone a formal CMMI appraisal, its practices exhibit characteristics associated with the higher maturity levels—in particular, an orientation toward continuous, feedback-driven improvement of the kind CMMI describes at its "Optimizing" level. Within this setting, the approach developed and deployed in this project added a capability that the sprint cadence did not previously provide: a systematic feedback loop that learns from the organization's accumulated history of past defects and the human errors behind them, and channels that learning back into process improvement, quality assurance, and engineer development. In other words, the approach complements the forward-looking, iteration-based rhythm of Agile with a backward-looking, evidence-based learning mechanism grounded in the organization's own defect and defect-fix data.

The approach is in active use in the organization. The classifications and periodic analytics have been incorporated into ongoing practice rather than remaining a one-off study, and the quality-assurance and capability-building elements described in Section 3.5 continue to operate. Consistent with the "improving" and "release" stages of the technology-transfer model adopted in Section 2.1, ownership of the approach has moved from a study conducted with external involvement toward a practice maintained by the organization itself. We are, moreover, already exploring ways to improve the approach—refining the classification schemes, extending the analyses, and improving how the AI assistance and the periodic feedback are integrated into the development workflow—which we outline as future work in Section 6.

## 5 DISCUSSIONS

We now step back from the individual results to consider what the study means: the lessons it yielded, its implications for researchers and practitioners, how far its findings can be possibly generalized, and the limitations that qualify them.

### 5.1 Lessons learned

Several lessons emerged from developing and deploying the approach. The first is that the value lay less in any single analytic technique than in the discipline of systematically learning from data the organization already held. The defect and defect-fix records had accumulated for years and were used mainly to close individual issues; the contribution was to turn this under-used history into a basis for prevention and learning. A related lesson is that the data's usefulness depended on the quality of what engineers recorded, and that establishing a culture of precise, technical root-cause and fix descriptions was as important as the analysis itself.

A second lesson concerns granularity. Classifying defect origins only at the level of coarse SDLC phases proved too blunt to act on; the finer-grained scheme, which localized origins to back-end logic, database code, and specific data-related activities, was what made the findings actionable. A third lesson is that AI assistance was genuinely useful for the labor-intensive parts of the analysis—classification, aggregation, and synthesis—but only under consistent human oversight; treating AI output as a candidate for expert review, rather than as an answer, was essential to trustworthy results. Finally, the individual-feedback element taught us that engineer-level analysis can support learning without becoming evaluation, but only if it is deliberately framed and safeguarded; the framing was not incidental but central to its acceptance.

### 5.2 Implications for Researchers

For researchers, the study offers a concrete instance of how generative AI can be integrated into defect analytics under human governance, and how such an approach can be evaluated in a real industrial setting rather than in the abstract. It suggests that the analysis of defect-fix data—not merely defects—is an under-explored source of insight into prevention, and that human-error origins, examined at a fine granularity, connect defect analytics to the wider literature on human error and cognitive models in software engineering [14-17]. The study also illustrates a pragmatic evaluation design—triangulating practitioner judgement, classification validation, and an effort-normalized pre/post analysis—for settings where controlled experimentation is not feasible, and points to the effort-normalized measurement of defect density (per task and per actual hour) as an alternative to KLOC-based measures where much work modifies existing code. Each of these is a direction that further, more rigorous research could develop.

### 5.3 Implications for Practitioners

For practitioners, the study shows that an organization does not need advanced tooling or large investment to begin learning systematically from its own defects: the raw material is typically already present in issue-tracking systems, and much of the analytical heavy lifting can be delegated to AI assistance under expert review. The approach is copyable in outline—capture root-cause and fix information consistently, classify defects by both failure and origin, analyze recurring patterns, and feed the results into process, assurance, and capability improvements. The study also offers a cautionary, practical lesson: engineer-level defect data is sensitive, and its use for learning rather than evaluation depends on explicit safeguards and framing, without which the same analysis could damage trust rather than build capability.

### 5.4 Generalizability of the Findings

The study concerns a single organization, and its specific quantitative findings—the distribution of defect origins, the accuracy figures, and the observed change in defect density—should not be assumed to transfer directly to other contexts, which differ in technology, domain, process maturity, and team composition. What is more likely to generalize is the *approach* and the *problem it addresses*. As argued in Section 2.4, many organizations accumulate substantial defect and defect-fix data while making limited systematic use of it, and the method described here—AI-assisted, human-governed analysis of that data for prevention, assurance, and learning—is not specific to the studied organization's systems. The finer-grained, origin-level view of defects, and the practice of learning from fixes rather than only from failures, are likewise transferable in principle. We therefore present the approach as an instance offered for adaptation rather than as a solution whose results will replicate unchanged, consistent with the technology-transfer framing of Section 2.1.

### 5.5 Limitations and Threats to Validity

Given the empirical nature of this research, the findings should be interpreted in light of several limitations and potential threats to validity. Like any empirical study, the investigation reported in Section 4 is subject to constraints related to context, methodology, and data sources. In the following, we discuss these using four commonly adopted categories of validity threats [32]—internal validity, construct validity, conclusion validity, and external validity—and outline the steps taken to mitigate them.

*Internal validity* concerns whether the observed improvements can be attributed to the approach. This is the study's principal limitation. The evaluation is a single-group, pre/post design without a control group, and the periodic feedback was introduced alongside the process and quality-assurance changes of Section 3.5. Any improvement in defect density (RQ3) cannot be attributed to the feedback, nor isolated from concurrent influences such as growing engineer experience, increasing maturity of the systems themselves, changes in workload or testing intensity, the parallel process and quality-assurance changes, or the effect of being observed. We have not claimed causation, and we report RQ3 as indicative rather than conclusive; we also probed attribution directly in interviews, which pointed to a combination of factors rather than any single cause. We therefore characterize the RQ3 finding as a tentative temporal association between the introduction of the approach and a subsequent reduction in defect density, rather than as evidence of a causal effect, and we present the per-developer distribution (Figure 8) rather than aggregate means alone so that the variation underlying this association is visible.

*Construct validity* concerns whether our measures capture what they intend to. The defect classifications rest partly on human judgement, and our reliability checks, while positive, were limited: the joint review was not blind, and the independent blind re-classification, though stronger, covered a smaller sample. Defect density is measured over *reported* defects rather than all defects, and is normalized by tasks and actual hours rather than code size; these are deliberate choices (Section 4.6.3) but they mean the metric reflects escaped-defect density per unit of effort, not total defect density. The developer-level error attribution (Analysis 4), unlike the classifications, was reviewed by the research team but not subjected to an independent validation; it also involves inference over version history and is inherently uncertain in some cases, particularly for defects with more than one contributing error. Independently validating this attribution is left to future work.

A further threat, to both reproducibility and construct validity, concerns the specification of the AI models used. As noted in Section 3.4.1.1, the analyses were performed with different models over the study period—initially GPT-4o and subsequently successive versions of Claude Sonnet—and we selected the latest available version at each point rather than fixing a single model version throughout. We did not record the exact model version, decoding parameters, or context-management settings in use for each analysis. This limits the exact reproducibility of the AI-assisted steps: another team repeating the work with a different model or version could obtain somewhat different classifications or suggestions, and we cannot fully attribute our results to a specific, frozen model configuration. Two considerations partly mitigate this threat. First, the AI was used throughout as an aid whose outputs were reviewed and validated by human experts (Section 3.4.1.1), so the classifications and recommendations ultimately reflect human judgement rather than any single model's raw output. Second, as reported in Section 3.4.1.1, we did not observe major differences in output quality across the model versions used for our specific tasks, which suggests the approach is not tightly coupled to one particular model. Nonetheless, we recommend that future applications and replications record the exact model version, parameters, and configuration used, as we intend to do in subsequent work.

*Conclusion validity* concerns whether our inferences from the data are sound. Given the sample sizes—374 defects, 19 survey respondents, 19 developers in the density analysis, and a 71-defect validation sample—we deliberately report descriptive statistics and agreement rates rather than inferential tests, and we avoid over-interpreting small differences. The self-reported survey items on behavior change are subject to acquiescence and social-desirability bias, which is why they are treated as corroboration only, with the quantitative density analysis carrying the main weight for RQ3.

*External validity* concerns generalization, discussed in Section 5.4. The single-organization, single-context nature of the study limits statistical generalization; we mitigate this by describing the approach in transferable terms and by being explicit about what is context-specific.

The approach as applied relied on cloud-based commercial AI services, to which anonymized defect data—and, for the attribution step, source-code content via the company's existing Claude Code–Git integration—were transmitted. Direct identifiers were removed beforehand and a commercial license excluding training use was used, but anonymization of code relied on an instructed engineering practice rather than a technical guarantee, and no zero-retention or on-premise deployment was used; organizations with stricter requirements may need such measures.

Finally, a limitation cutting across these categories is the researcher's dual role. Consistent with the action-research setting (Section 2.1), the author participated in developing, deploying, and evaluating the approach, and the senior technical managers who validated classifications were also involved in defining them, and hold the anonymization key (Section 3.6). We mitigated this through the independent blind re-classification, the anonymous practitioner survey, and the documented adoption evidence, none of which depends solely on the research team's own assessment; but we acknowledge it as a genuine constraint on the independence of the evaluation.

## 6 Conclusions and Future Work

This paper has presented an AI-assisted approach for analyzing an organization's historical defect and defect-fix data to support defect prevention, and reported an industrial case study of its use. The approach classifies defects by both their observed failure and their human-error origin, examines how each could have been prevented, and synthesizes the results into feedback for process improvement, quality assurance, and engineer capability building—all under consistent human oversight of the AI's outputs. Evaluated through a triangulated case study, the approach produced classifications that practitioners and independent review judged accurate, outputs that practitioners found useful and actionable, and a decrease in effort-normalized defect density following the introduction of feedback. Because the evaluation is an uncontrolled, single-group pre/post comparison, we interpret this as a tentative temporal association rather than evidence of a causal effect, since the change cannot be separated from concurrent factors such as growing system maturity and the parallel process and quality-assurance improvements.

Distinct from prior work centered on defect detection and prediction, the approach's contribution lies in learning systematically from defect *fixes* as well as defects, at a fine origin-level granularity, and in integrating this learning into an organization's routine practice in a human-governed way.

Several directions remain for future work. The approach itself can be refined—sharpening the classification schemes, extending the analyses, and improving how AI assistance and periodic feedback are integrated into the development workflow. Methodologically, stronger evaluation designs, including comparison across teams or a longer longitudinal study with clearer separation of the feedback from concurrent changes, would allow firmer claims about impact. Finally, applying the approach in other organizations would test its transferability and help distinguish which findings are context-specific from those that hold more widely.

## DECLARATION STATEMENTS

**Data Availability Statement.** The dataset analyzed in this study consists of proprietary defect and defect-fix records from the participating company and contains commercially and personally sensitive information; it therefore cannot be shared publicly. The survey and interview instruments are provided in Appendices A and B, and the classification schemes and representative prompts are described in Section 3.4, to support methodological transparency and partial replication.

**Declaration of Competing Interest.** The author served as a consultant to the company in which this study was conducted, in the course of the action-research engagement described in the paper. This relationship is inherent to the industrial, action-research nature of the work. The author received no financial incentive contingent on the study's outcomes, and the company placed no restrictions on the reporting of negative or unfavorable findings.

**Declaration of ethics considerations for the survey and interviews.** The survey and interview study did not undergo formal institutional review board (IRB) approval, as it was conducted within an industrial setting where such review is not routinely required. Nonetheless, it was carried out in accordance with standard research-ethics principles: participation was voluntary and could be withdrawn without consequence, the survey was anonymous, interviews were recorded and transcribed only with consent, and all individual-level data were anonymized and handled under the safeguards described in Sections 3.4.5 and 3.6.

**Declaration on generative-AI use in drafting the manuscript text.** During the preparation of this manuscript, the author used the Claude GenAI tool with the "Opus 4.8" LLM to assist with drafting, editing, and improving the language of the text. The author reviewed and edited all such content and takes full responsibility for the content of the publication. Generative AI's use as an analytical instrument within the study itself is described as part of the methodology in Section 3.4.1 and is distinct from this declaration.

## ACKNOWLEDGEMENTS

The author would like to thank the management and software engineering teams of the case-study company for their active involvement in this R&D project.

## Appendix A — Practitioner Survey Instrument

*Introductory note shown to respondents:* This survey is part of a study of the AI-assisted defect and defect-fix analytics approach used in our teams. Its purpose is to learn how useful and accurate the approach is, so that it can be improved. Participation is voluntary and responses are anonymous; there are no right or wrong answers, and your candid views—including criticism—are valued. You may skip any question or withdraw at any time.

*Respondent background (non-identifying)*

1. Which best describes your main role? ☐ Developer ☐ QA/Test ☐ Technical lead / senior technical manager ☐ Other
2. Approximate years of software engineering experience: ☐ <2 ☐ 2–5 ☐ 6–10 ☐ >10

*Part A – Accuracy of the classifications (RQ1).* For the defects associated with your own work:

3. The failure-type classifications assigned to my defects were accurate. (Strongly disagree – Strongly agree, 5-point)
4. The identified origins (the human error / SDLC phase behind my defects) were accurate. (5-point)
5. *(Free text)* If you disagreed with any classification of your defects, please give an example and explain.

*Part B – Usefulness and actionability (RQ2)*

6. The analytics gave a clearer picture of where and how defects arise than we had before. (5-point)
7. The suggested process improvements were relevant and actionable. (5-point)
8. The suggested quality-assurance changes (e.g., earlier or more targeted reviews and checks) were relevant and actionable. (5-point)
9. The periodic feedback on my own recurring error patterns was accurate and fair. (5-point)
10. The periodic feedback helped me reflect on, and pay more attention to, specific aspects of my work. (5-point)
11. *(Free text)* Which part of the approach (defect analytics, process changes, QA changes, or individual feedback) was most useful to you, and why?
12. *(Free text)* What was least useful, unclear, or in need of improvement?

*Part C – Perceived impact (supports RQ3)*

13. Since receiving feedback, I have changed how I approach the kinds of tasks where my defects tended to occur. (5-point)
14. I have noticed the error types highlighted for me recurring less often in my recent work. (5-point)
15. *(Free text, optional)* If you have observed any change in your own work as a result of the approach, please describe it.

*Part D – Overall and ethics*

16. Overall, the approach is a worthwhile use of the team's effort. (5-point)
17. The way individual feedback was handled (private, for learning rather than evaluation) was appropriate. (5-point)
18. *(Free text, optional)* Any other comments, including any concerns about how the approach or the data were used?

## Appendix B — Semi-Structured Interview Guide

*Format:* approximately 45 minutes, audio-recorded and transcribed with consent; participants spanning developers, QA staff, and senior technical managers. The guide below lists themes and lead questions; follow-up probes were used as appropriate, and participants were free to raise issues not listed.

*Opening.* Briefly, what has been your role in relation to the defect analytics approach, and how much contact have you had with its outputs?

*Theme 1 – Accuracy and trust (RQ1).* How accurate did you find the classifications of defects—both the type of failure and the origin/human error behind it? Where did they get it right, and where did they get it wrong? How much did you trust the AI-assisted parts, and what made you trust or doubt them?

*Theme 2 – Usefulness and actionability (RQ2).* Which outputs of the approach did you actually use, and for what? Were the insights about where defects originate new to you, or already known? Which of the derived changes—process improvements, quality-assurance changes, or individual feedback—made a real difference in practice, and which did not? Can you give a concrete example of something you did differently because of the approach?

*Theme 3 – Individual feedback and its framing (RQ2, ethics).* How did the periodic feedback on recurring error patterns land for you (or, for managers, for your team)? Was it accurate and fair? Did it feel developmental or evaluative? Was the private, learning-oriented framing appropriate, and would you change anything about how it is handled?

*Theme 4 – Impact over time (RQ3).* Over the months after the feedback began, did you notice any change in the kinds of defects arising in your own work, or your team's? What do you attribute that to—the feedback specifically, the process and QA changes, growing experience, or other factors? (Probe for attribution, to surface confounds.)

*Theme 5 – Adoption and sustainability (Section 4.7).* Which parts of the approach are still in use now, and which faded? What made some changes stick and others not? What would be needed for this to continue without external involvement?

*Closing.* If you could change one thing about the approach, what would it be? Anything else you would like to add?

## Appendix C —Definitions and Examples of the Classification Categories

We shall note that each defect was assigned to exactly one category, using the precedence rule described in Section 3.4.2.

**Table C.1- The eleven failure-type categories (Analysis 1)**

| # | Failure category | One-sentence definition | Example |
|---|---|---|---|
| 1 | Unhandled Exception or Crash | The system raises an unhandled runtime error or terminates unexpectedly instead of completing the action. | A page returns HTTP 500 when saving a record. |
| 2 | Incorrect Computation & Business Logic | The system runs and returns a result, but the computed value or applied rule is wrong. | A salary calculation applies the wrong allowance and returns an incorrect total. |
| 3 | Incorrect Data Display & Formatting | Data is retrieved correctly but shown wrongly—wrong value, wrong field, or wrong format. | A report displays a value in the wrong column or with the wrong number format. |
| 4 | Data Persistence & Loss | Data that should be saved is not persisted, is partially saved, or is lost. | A "Save and close" action completes but the entered changes are not stored. |
| 5 | Data Retrieval & Search | Stored data cannot be found, searched, or returned when it should be. | A search for a newly added record returns "no data available." |
| 6 | Data Consistency & Synchronization | The same data differs across screens, reports, or integrated modules, or is duplicated. | A total shown on one screen differs from the same total on a report. |
| 7 | Function Not Executing | A feature, button, or action does not trigger or produces no effect at all. | Clicking "Copy document" performs no action. |
| 8 | Input Validation & Handling | The system accepts input it should reject, rejects valid input, or gives a wrong/missing validation response. | A mandatory field is left empty but the system raises a database error instead of a validation message. |
| 9 | Workflow & State Transition | A multi-step process advances to a wrong step, skips a required stage, or blocks a valid transition. | An approval step is reached before a required review stage is completed. |
| 10 | Authorization & Access Control | A user can access what they should not, or cannot access what they should. | A user is denied access to a resource their role permits. |
| 11 | Performance & Responsiveness | The behaviour is correct but too slow—timeouts, freezing, or slow response under load. | A list view times out when loading under normal data volumes. |

**Table C.2 — The eighteen origin (human-error) categories (Analysis 2), grouped by SDLC phase.**

| # | SDLC phase | Origin category | One-sentence definition | Example |
|---|---|---|---|---|
| 1 | **Requirements** | Requirements analysis error | A defect caused by a misunderstood, incomplete, or incorrectly specified requirement. | An eligibility rule specified incorrectly leads to a wrong calculation. |
| 2 | **Design** | Database design error | A defect originating in the data model, such as a wrong table structure or relationship. | A missing relationship causes a query to return incorrect joined data. |
| 3 | | UI design (non-front-end-code) error | A defect arising from an incorrect user-interface design decision, independent of its coding. | A mandatory field is shown as optional in the designed form. |
| 4 | **Coding** | Coding error — front-end | A defect introduced in client-side code. | An incorrect parameter is sent from the client to the server. |
| 5 | | Coding error — back-end controller | A defect introduced in server-side control/business logic. | Save logic relies on an inappropriate identifier field. |
| 6 | | Coding error — back-end T-SQL | A defect introduced in database (T-SQL) code such as queries or stored procedures. | A stored procedure omits a component, producing a wrong computed value. |
| 7 | **Deployment, configuration, installation** | Code deployment error | A defect introduced by an incorrect or mis-sequenced deployment step. | A correct build fails because a deployment step was skipped. |
| 8 | | Installation error — initial data | A defect caused by missing or incorrect reference/seed data at installation. | A required status row is absent, so a lookup fails. |
| 9 | | Installation error — server/tools | A defect caused by a missing or misinstalled server component or tool. | A database engine is installed without a required full-text search component. |
| 10 | | Configuration error | A defect caused by an incorrect configuration setting, with correct code. | A display flag is initialized to the wrong value, hiding a screen element. |
| 11 | | Other installation error | An installation-related defect not covered by the categories above. | — |
| 12 | **Software maintenance** | Data change-management error | A defect caused by an incorrectly applied schema change, migration, or production-data update. | A duplicated data change causes records to be counted twice. |
| 13 | | Regression (fix breaks another part) | A defect introduced while fixing another defect. | A correction to one module disables a previously working feature. |
| 14 | **Ungrouped** | Test error | The reported "defect" is an error in the test case, script, or expected result rather than in the system. | A test asserts an incorrect expected value and fails a correct system. |
| 15 | | End-user error | The reported issue stems from incorrect use rather than a system fault. | A user follows an unsupported sequence and reports the outcome as a bug. |
| 16 | | Defects by other external systems | The defect originates in an external system the team does not control. | An integrated third-party service returns malformed data. |
| 17 | | Bug cannot be reproduced | The reported defect could not be reproduced and no origin could be established. | An intermittent report that did not recur under investigation. |
| 18 | | Not a bug | The report was determined not to correspond to a genuine defect. | A request for a feature change filed as a bug. |

## APPENDIX D — REPRESENTATIVE PROMPTS USED IN THE AI-ASSISTED ANALYSES

The prompts below are a representative, anonymized subset of those used across the analyses. They are not an exhaustive log: prompts were refined pragmatically and iteratively during the work, and some contained company-, project-, or developer-specific information that cannot be disclosed for confidentiality reasons. The examples here have been lightly edited to remove identifying detail (shown as bracketed placeholders such as [system], [project], [DEVx]) while preserving the structure and intent of the original prompts. They are grouped by the purpose they served.

### D.1 Establishing and agreeing the classification schemes

- "*For the new failure-classification column, propose a useful category set that will help us decide which QA activities to strengthen.*"
- "*Review this draft list of failure categories and point out any overlaps or ambiguous boundaries between them; suggest clearer wording where two categories could capture the same defect.*"
- "*Propose a set of origin categories that map each defect to the human error and SDLC phase where it was introduced, finer-grained than the standard requirements/design/implementation/testing phases.*"
- "*Before you start classifying, confirm that we are using this final set of categories (no overlap, clear wording), then classify this sample of [n] records so I can check the logic before applying it to the whole sheet.*"
- "*For each category in the agreed set, write a one-sentence definition and give one short example, so we can use them consistently.*"

### D.2 Analysis 1 — Failure-type classification

- "*Classify every row in this file using the agreed failure categories, preserving the exact bug text for each row.*"
- "*Go through the [column] bug descriptions in order and assign each one to the agreed failure categories (failure level, not human-error level).*"
- "*For any record you are unsure about, mark it as uncertain and briefly say which two categories it falls between, rather than guessing.*"
- "*Apply the precedence rule: if the system crashed or raised a runtime error, classify as Unhandled Exception; else if nothing happened, Function Not Executing; otherwise the specific wrong-behaviour category.*"
- "*Summarize the distribution across the failure categories for this batch, with counts per category.*"
- "*Re-check these [n] records I have flagged, and tell me whether you agree with the original failure classification or would change it, with a one-line reason.*"

### D.3 Analysis 2 — Origin (human-error) classification

- "*Using the short defect description, the root cause, and how the defect was resolved, identify the human error that led to each defect and the SDLC phase in which it occurred.*"
- "*For records where the origin field is empty, infer the most likely origin category from the root-cause and fix text, and mark it as AI-inferred so it can be validated separately.*"
- "*Where a defect appears to have more than one contributing human error, list each origin separately rather than forcing a single category.*"
- "*For this batch, distinguish coding origins into front-end, back-end controller, and back-end database (T-SQL), based on what the fix actually changed.*"
- "*Summarize the distribution of origins across the SDLC bands (design, coding, deployment/configuration/installation, maintenance) for these records.*"
- "*Flag any record where the observed failure and the inferred origin seem inconsistent, so I can review it.*"

## D.4 Analysis 3 — Prevention recommendations

- *"For each defect, suggest a specific quality-assurance or development practice that, applied at the right point, would plausibly have prevented it — tie the recommendation to the record's own root cause and fix."*
- *"For these back-end coding defects, explain how each could have been prevented, tying the recommendation to the failure and the human-error type captured for that row."*
- *"Group these prevention recommendations by origin category and identify the recommendations that recur, so we can prioritize the most impactful controls."*
- *"For the data-related origins (database design, initial-data installation, data-change management), propose preventive controls that apply earlier than testing."*
- *"Keep each recommendation concrete and actionable; avoid generic advice such as 'test more' or 'review code'."*

## D.5 Analysis 4 — Engineer attribution and per-engineer synthesis

Engineer attribution (via version-history analysis):

- *"Examine the change history and related files for the code associated with this defect in [repository], and reconstruct the sequence of changes that led to the introduced fault; identify which change(s) introduced it."*
- *"Where a defect's fault was introduced across changes by more than one contributor, list all contributors whose changes were involved rather than attributing it to one."*
- *"Cross-check the resolving engineer recorded for this defect against the change history, and note where the engineer who introduced the fault differs from the one who fixed it."*

Per-engineer pattern synthesis:

- *"For [DEVx], summarize the recurring error types across their associated defects, and identify the one or two areas where their defects most often concentrate."*
- *"Compare [DEVx]'s recurring pattern this period with the previous period, and note whether the flagged error type is recurring less often."*
- *"For each developer in [project], produce a short profile of their most frequent recurring error area, drawing only on that developer's own defects."*

## D.6 Generating the periodic developer feedback

- *"Draft a short, private feedback message for [DEVx] summarizing the most common recurring pattern in their own recent defects, with one or two concrete examples from their own work and a specific, forward-looking suggestion for improvement."*
- *"Keep the tone developmental and non-evaluative; frame it as a prompt for reflection, not an assessment, and include no comparison to other developers."*
- *"Include, where relevant, the preventive practice identified for that error type, as a starting point the developer can act on."*
- *"Revise this draft so it refers only to the developer's own defects and contains no ranking, score, or team-relative figure."*